\documentclass[acmsmall,pdftex,prologue,table,dvipsnames]{acmart}

\setcopyright{acmcopyright}
\copyrightyear{20??}
\acmYear{20??}
\acmDOI{??}

\acmJournal{TOMS}
\acmVolume{??}
\acmNumber{?}
\acmArticle{???}

\usepackage{wrapfig}
\usepackage{array}
\usepackage{textcomp}
\usepackage[T1]{fontenc}
\usepackage[utf8]{inputenc}	
\usepackage{stmaryrd}
\usepackage{color}     
\usepackage{xspace}
\usepackage{subcaption}
\usepackage{multirow}
\usepackage{makecell}

\defcitealias{OpenMP25}{OpenMP ARB, 2005}
\defcitealias{OpenMP40}{OpenMP ARB, 2013}
\defcitealias{OpenMP50}{OpenMP ARB, 2018}
\defcitealias{OpenMP51}{OpenMP ARB, 2020}
\defcitealias{OpenMP52}{OpenMP ARB, 2021}

\usepackage{verbatim}

\usepackage{listings}
\lstdefinestyle{customcpp}{
  breaklines=true,
  xleftmargin=\parindent,
  language=C++,
  showstringspaces=false,
  basicstyle=\small\ttfamily,
  keywordstyle=\bfseries\color{green!40!black},
  commentstyle=\itshape\color{purple!40!black},
  identifierstyle=\color{blue},
  stringstyle=\color{orange},
  tabsize=2,
  escapeinside={/@}{@/},
  keepspaces=true,
  captionpos=b,
  deletekeywords={for,private},
  morekeywords={[2]omp,parallel,sections,single,critical,ordered,barrier,section,master,dummy},
  morekeywords={[2]directive,firstprivate,private,shared,schedule,guided,declare,reduction,initializer,omp_priv,dummy},
  morekeywords={[2]omp_out,omp_in,for,num_threads,flush,atomic,nowait,copyprivate,dummy},
  morekeywords={[2]read,write,update},
  keywordstyle={[2]\bfseries\color{blue!40!black}},
  numbers=left,
}
\newcommand\lsttable[2][]{\lstinline[#1]{#2}}

\newlength\figureheight 
\newlength\figurewidth 

\newcommand{\aeq}{\kern.35em\text{\small+}\kern-.35em=}

\newcommand{\RR}{\mathbb{R}}

\newcommand{\intd}{\,\mathrm{d}}
\newcommand{\T}{\mathrm{T}}

\newcommand{\cpp}{C\texttt{++}\xspace}

\title{Event-Based Automatic Differentiation of OpenMP with OpDiLib}
\author{Johannes Blühdorn}
\email{johannes.bluehdorn@scicomp.uni-kl.de}
\author{Max Sagebaum}
\author{Nicolas R. Gauger}
\affiliation{%
	\institution{Chair for Scientific Computing, Technische Universität Kaiserslautern}
	\city{Kaiserslautern}
	\country{Germany}
}

\begin{document}

\begin{abstract}
We present the new software OpDiLib, a universal add-on for classical operator overloading AD tools that enables the automatic differentiation (AD) of OpenMP parallelized code. With it, we establish support for OpenMP features in a reverse mode operator overloading AD tool to an extent that was previously only reported on in source transformation tools. We achieve this with an event-based implementation ansatz that is unprecedented in AD. Combined with modern OpenMP features around OMPT, we demonstrate how it can be used to achieve differentiation without any additional modifications of the source code; neither do we impose a priori restrictions on the data access patterns, which makes OpDiLib highly applicable. For further performance optimizations, restrictions like atomic updates on adjoint variables can be lifted in a fine-grained manner. OpDiLib can also be applied in a semi-automatic fashion via a macro interface, which supports compilers that do not implement OMPT. We demonstrate the applicability of OpDiLib for a pure operator overloading approach in a hybrid parallel environment. We quantify the cost of atomic updates on adjoint variables and showcase the speedup and scaling that can be achieved with the different configurations of OpDiLib in both the forward and the reverse pass.
\end{abstract}

\ccsdesc[500]{Mathematics of computing~Automatic differentiation}
\ccsdesc[500]{Computing methodologies~Shared memory algorithms}
\ccsdesc[500]{Mathematics of computing~Mathematical software performance}
\ccsdesc[300]{Software and its engineering~Software design engineering}

\keywords{Algorithmic Differentiation, OpenMP, OMPT, High Performance Computing, AD Tool Design, Reusable Software}

\maketitle

\section{Introduction}
\label{section:introduction}

Automatic differentiation (AD) deals with the evaluation of derivatives of computer codes, in a machine accurate manner and for a computational cost that is often only a small multiple of the cost of the codes themselves. AD has repeatedly proven its value in scientific computing, especially for large simulation codes where symbolic derivatives are not feasible and finite differences produce numerically unstable and unreliable results. Areas of application include discrete adjoint solvers \cite{SchlenkrichWGH2008,TowaraN2013}, shape optimization \cite{OezkayaG2010,VerstraeteMM2017,VitalePC2020}, inverse problems \cite{AurouxG2017}, and machine learning \cite{BaydinPRS2017,GuentherRSCG2020}. A comprehensive introduction to AD is given in \cite{GriewankW2008}.

We view a computer program with input variables $x$ and output variables $y$ as a mathematical function $F\colon\RR^n\to\RR^m\colon x\mapsto y$. While an execution of this program for a specific input $x$ might involve millions of code statements, it always amounts to a sequence of \emph{elementary operations} like sums, products, or standard math library function like $\sin$ or $\exp$. For all these operations, the derivatives are known; hence, we can evaluate the derivative of the full code with respect to $x$ by a rigorous application of the chain rule. The \emph{reverse mode of AD} is a specific strategy for the structured traversal of the chain rule. For each variable $v$ of the code, whether input, output, or intermediate, we introduce an \emph{adjoint variable} $\bar{v}$ that fulfills the relation $\bar{v}=\frac{\intd F}{\intd v}(v)^\T\bar{y}$. For the adjoint inputs $\bar{x}$, we obtain
\begin{equation*}
\bar{x}=\frac{\intd F}{\intd x}(x)^\T\bar{y}
\end{equation*}
so that suitable choices of $\bar{y}$ yield various derivatives of $F$, and hence the computer program at hand. Choosing $\bar{y}$ is known as \emph{seeding}. In an implementation, it is practical to assemble the adjoint variables in an incremental fashion statement by statement. Let a code statement be given by
\begin{equation}
\label{eq:forward_statement}
w=\phi(u)
\end{equation}
where $w\in\RR$ is an intermediate variable or an output, $u\in\RR^k$ consists of inputs and intermediate variables from previous statements, and $\phi\colon\RR^k\to\RR$ is composed of one or multiple elementary operations. The adjoint updates for the statement read
\begin{equation}
\label{eq:adjoint_updates}
\bar{u}_j\aeq\frac{\partial\phi}{\partial u_j}(u)\bar{w}\hspace{0.2cm}\text{ for }j=1,\,\ldots,\,k,\text{ then set }\bar{w}=0.
\end{equation}
Compared to the data flow implied by a sequence of statements \eqref{eq:forward_statement}, the data flow of the corresponding adjoint updates \eqref{eq:adjoint_updates} is reversed. Derivative information is propagated from the adjoint outputs $\bar{y}$ via adjoint intermediate variables to the adjoint inputs $\bar{x}$. This data flow reversal challenge can, for example, be approached as follows. First, the code is executed in its original structure in a \emph{forward pass}; throughout, it is \emph{recorded} in a suitable representation on a stack-like structure called \emph{tape}. The tape stores all information required for the reversal of all statements, that is, identification for the adjoint variables $\bar{u}$ and $\bar{w}$ \cite{SagebaumBG2020}, and information for the computation of the derivatives $\frac{\partial\phi}{\partial u_j}(u)$ \cite{Hogan2014,SagebaumAG2018}. The order in which all statements \eqref{eq:forward_statement} were executed in the forward pass is preserved in their recording on the tape. In a subsequent \emph{reverse pass}, the tape is \emph{evaluated}, that is, it is unwound from end to start to provide the data for the corresponding adjoint updates \eqref{eq:adjoint_updates} in reverse order.

A key objective of \emph{AD tools} is the automatic augmentation of a given source code by extra logic required for the computation of derivatives. The two main approaches for this are \emph{source transformation} and \emph{operator overloading}. Source transformation tools like Tapenade \cite{HascoetP2013} are essentially compilers that take the given source code and transform it directly to the augmented source code. Operator overloading tools rely on a replacement of the floating point data type by a custom \emph{AD type}. All elementary operations are overloaded for the AD type. From the user's point of view, the AD type behaves like a floating point type; internally, the overloads execute the extra logic that is required for differentiation. Operator overloading tools usually implement the taping approach described above, examples include ADOL-C \cite{Walther2009}, Adept \cite{Hogan2014}, and CoDiPack\footnote{\url{https://www.scicomp.uni-kl.de/software/codi/}} \cite{SagebaumAG2019}.

In order to achieve maximum performance on today's multicore architectures, simulation codes typically employ shared memory parallelism, for example by means of OpenMP \citepalias{OpenMP52}. Differentiating such codes requires AD tools that can handle OpenMP directives. Research on combining AD and OpenMP started in the realm of source transformation tools when \cite{BueckerLMB2001,BueckerLRBM2002} suggested the generation of OpenMP parallel code for the simultaneous computation of multiple derivatives of an otherwise serial code with the forward vector mode. \cite{BueckerRW2004} extended this to the case of an already parallel code by using nested parallelism. First reports on source-to-source reverse mode differentiation of an OpenMP parallel simulation code are given in \cite{HeimbachHG2002,HeimbachHG2005}. Subsequently, support for reverse mode AD for OpenMP parallel codes was established in operator overloading tools as well, but faced the challenge that OpenMP parallelism is specified in large parts via \lstinline{#pragma} directives. Techniques like operator overloading do not apply, and unlike, e.\,g., an MPI function call, an OpenMP directive may influence the control flow beyond its immediate location in the code. For a parallel region, e.\,g., the compiler will introduce thread handling both before and after the contained code. Hence, it is much harder to change the directives' behaviour without rewriting the source code. ADOL-C relies on the side effects of constructors and destructors of privatized objects to establish OpenMP support for parallel regions and parallel loops \cite{Kowarz2008,BischofGKW2008}. The source code rewriting can be reduced to a manual annotation of the respective pragma directives for these OpenMP features, as opposed to, e.\,g., annotations both at the beginning and end of parallel regions. Recently, it was shown in conjunction with Adept how compiler instrumentation can be used to introduce extra logic for the differentiation of general recursive fork-join programs in a fully automatic fashion \cite{KalerSXLCPK2021}. Source transformation tools, on the other hand, have unlimited access to OpenMP directives. Transformation rules for many additional OpenMP directives were established in \cite{Foerster2014}. Recently, capabilities to differentiate OpenMP parallel worksharing loops have been added to Tapenade, with a particular focus on the appropriate treatment of variable scopes and execution schedules \cite{HueckelheimH2021}.

We summarize some key insights from previous research. Transitioning from the reverse mode of AD in a serial context as described above to parallel codes, we face the problem that taping is an inherently sequential process. Therefore, computations that are performed simultaneously in different threads have to be recorded on different tapes. Furthermore, it was observed that the parallelization of the forward pass can be used to deduce a corresponding parallelism for the reverse pass. However, this is --- except for special cases --- not as simple as evaluating the tapes in parallel because any additional synchronization that occurred in the forward pass has to be transformed to corresponding reverse synchronization. Furthermore, while simultaneous read access to the same variable from different threads is fine in the forward pass, the corresponding adjoint updates \eqref{eq:adjoint_updates} result in simultaneous write access to the same adjoint variable, which introduces data races in the reverse pass. This can be addressed conservatively by performing all adjoint updates atomically \cite{Foerster2014}. In general, it is challenging to automatically identify variables that need atomic updates. Recently, an efficient parallel reversal approach was presented that does not rely on atomic updates \cite{KalerSXLCPK2021}. Instead, incremental adjoint contributions are stored in dedicated memory locations and their reduction is delayed until the reversal of the statement that wrote the corresponding primal variable, with the overhead of providing and managing these memory locations.

In this paper, we present a novel event-based approach for the reverse mode automatic differentiation of OpenMP parallel codes with operator overloading tools, and its implementation in the new open source tool OpDiLib\footnote{\url{https://www.scicomp.uni-kl.de/software/opdi/}}. Starting with version 5.0 of the OpenMP specification \citepalias{OpenMP50}, internal parts of the OpenMP runtime are formalized by \emph{execution model events}. Furthermore, OMPT becomes part of the specification, an API that allows tool developers to observe and react to these events \cite{EichenbergerEtAl2013}. OMPT was originally designed to provide a standardized interface for the performance monitoring of OpenMP applications; more generally, it allows for the coupling of the OpenMP runtime with tools that execute in the address space of the OpenMP parallel program \citepalias{OpenMP52}, which is precisely the case for operator overloading AD tools. In this work, we identify and demonstrate the applicability of OMPT for automatic differentiation. Specifically, it allows us to implement augmentations required for AD handling as side effects of OpenMP constructs, without any additional modifications of the OpenMP parallel source code.

In order to support OpenMP in an operator overloading AD tool, two kinds of adaptions must be made. First, the AD tool itself must be thread-safe. Second, an appropriate handling of OpenMP mechanisms must be provided. While certain OpenMP support has been crafted into some operator overloading AD tools by now, we demonstrate with OpDiLib for the first time that the second part can be implemented independently of a specific operator overloading AD tool. Moreover, this implementation is based on a limited set of features that many AD tools already have, which allows us to design OpDiLib as a universal add-on for classical operator overloading AD tools. Support for OpenMP in such a classical operator overloading tool is then achieved by making it thread-safe in a first step, and coupling it with OpDiLib in a second step. This continues a philosophy of AD tool agnostic support for parallelism in operator overloading that was established for MPI\footnote{\url{https://www.mpi-forum.org/}} in the tools AMPI \cite{SchanenNHU2010} and MeDiPack\footnote{\label{footnote:medipack}\url{https://www.scicomp.uni-kl.de/software/medi/}}. The coupling is achieved by implementing an AD tool wrapper, which we detail in Section \ref{section:ad_tool}. OpDiLib bindings are already implemented for CoDiPack. We hope that this paper is also a useful starting point for coupling OpDiLib with other operator overloading AD tools in the future.

OpDiLib's event-based differentiation logic is presented in Section \ref{section:logic}. It is formulated in terms of \emph{AD events} that mark the beginnings and ends of OpenMP constructs, and the AD handling is implemented as reactions to these events. This corresponds to an AD-centered point of view that allows for a unified handling of similar OpenMP mechanisms and hides many OpenMP and OMPT specific details that are not relevant for AD. Furthermore, the differentiation logic itself does not depend on OMPT at all. While OMPT execution model events can be used to generate AD events, the differentiation logic can also be reused together with other AD event generation mechanisms. With this approach, we establish in Section \ref{subsec:transformation_of_openmp} operator overloading AD support for the constructs of the OpenMP 2.5 specification \citepalias{OpenMP25}, with the exception of atomics and the flush directive, including several constructs that were only treated by source transformation tools previously.

From a user's point of view, OpenMP is very easy to apply. Introducing parallelism to a code can be as straightforward as supplying a loop with a \lstinline{#pragma omp parallel for} directive. To achieve maximum parallel performance, on the other hand, additional restructuring of the code and the application of more advanced OpenMP features might be required. OpDiLib follows the same philosophy. With OMPT, a first differentiated version of an OpenMP parallel code can be obtained without any additional code changes and without restrictions on the data access patterns. We achieve this by defaulting to atomic updates on adjoint variables. Starting from there, a user may employ additional optimizations. An important part of improving the performance of the parallel reverse pass is the elimination of atomic updates, which might only be possible after a revision of the data access patterns. Then, OpDiLib's adjoint access control tools which we present in Section \ref{subsec:data_access_patterns} can be used to disable atomic updates for selected parts of the code.

Section \ref{section:backend} presents two different backends for the generation of AD events. Section \ref{section:ompt_backend} details the generation of AD events via OMPT execution model events. OpDiLib is designed to be used with these modern OMPT features. Nonetheless, compilers are still in the process of establishing support for OMPT. As many projects rely on older compiler versions or compilers without OMPT features, we provide a second backend that does not require OMPT. It is based on an alternative interface to OpenMP functionality that consists of replacement macros for OpenMP \mbox{\lstinline{#pragma}} directives and clauses as well as replacements for OpenMP's runtime functions. The only additional step required for its application is the replacement of OpenMP constructs. As the macro interface is fully compatible with the OMPT backend, these code changes do not have to be revisited for a transition from the macro backend to the OMPT backend.
Section \ref{section:macro_backend} provides an overview over the macro interface and highlights some implementational details of the macro backend.

\begin{figure}
\centering
\includegraphics[width=\textwidth]{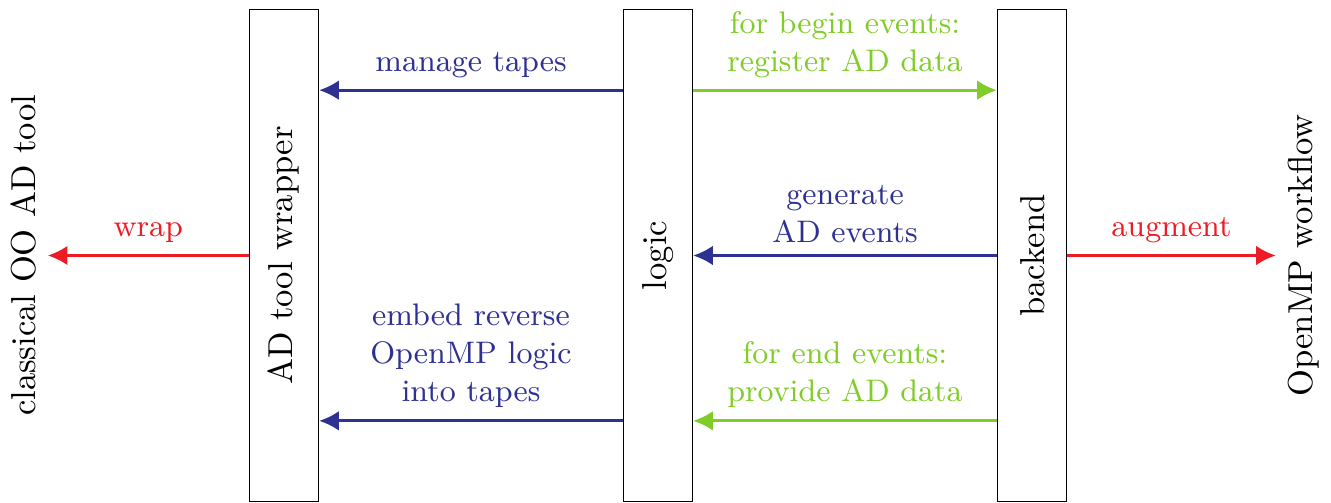}
\caption{Layers of OpDiLib's architecture. \textcolor{Blue}{\textbf{Blue arrows}} indicate important tasks of a layer for which it relies on functionality exposed by the pointed-to layer. Besides AD event generation, the backend layer is also responsible for AD data exchange between matching begin and end events. This AD data flow is indicated by \textcolor{LimeGreen}{\textbf{green arrows}}. OpDiLib builds a bridge between a classical operator overloading AD tool on the one hand and the OpenMP workflow on the other hand, the dependency relations to both indicated by \textcolor{Red}{\textbf{red arrows}}.}
\label{figure:opdilib_architecture}
\end{figure}

Together, the AD tool wrapper, the logic layer, and the backend layer make up OpDiLib's layered architecture. Figure \ref{figure:opdilib_architecture} visualizes the essential relations between these layers, with further details given in the respective Sections \ref{section:ad_tool}, \ref{section:logic}, and \ref{section:backend}.

In Section \ref{section:performance_study}, we conduct a performance study of OpDiLib applied to an OpenMP-MPI hybrid parallel code. We use CoDiPack as the underlying AD tool and MeDiPack for the differentiation of MPI. While a mixed approach between operator overloading and source transformation was established for hybrid parallel codes in \cite{SchanenFLLN2012}, we demonstrate with OpDiLib that they can also be handled by pure operator overloading, which admits a simplified toolchain without additional source transformation compilers and does not involve interfaces between different AD paradigms. We compare OpDiLib's backends, different compilers, quantify the cost of atomic updates on the adjoint vector and demonstrate the impact of OpDiLib on the parallel performance of the forward and reverse pass on two types of processors. We conclude our work in Section \ref{section:conclusion}.

\section{AD Tool Coupling}
\label{section:ad_tool}

In this section, we present the assumptions on the AD workflow that the implementation of OpDiLib is based on and collect the properties and features that an operator overloading AD tool must have in order to be coupled with OpDiLib. The coupling is achieved by an implementation of OpDiLib's AD tool wrapper interface which we showcase in this section. We close with a discussion of some advanced use cases.

To begin with, we assume that an AD workflow such as depicted in Listing \ref{listing:ad_workflow} is used, and that the AD tool supports the classical operations involved.
\lstset{basicstyle=\footnotesize\ttfamily}
\begin{lstlisting}[caption={AD workflow.}, label={listing:ad_workflow}]
// within serial code
AD::masterTape->registerInputs(inputs);
AD::masterTape->setActive(); // start recording
  
// function to be differentiated
// involves parallel code
f(inputs, outputs);  
  
AD::masterTape->setPassive(); // stop recording
AD::masterTape->registerOutputs(outputs);
AD::masterTape->seedOutputs(seeds);
AD::masterTape->evaluate(); // reverse pass
AD::masterTape->reset();
\end{lstlisting}
\lstset{basicstyle=\small\ttfamily}
In Listing \ref{listing:ad_workflow}, all parallel regions are assumed to be contained in \lstinline{f}.
Particularly, we assume that the recording is controlled by the initial implicit task, that is, serial parts of the code. 
Hence, when transitioning to parallel code, the general AD workflow set up for the serial version can be kept. As before, the user controls AD via the tape of the initial implicit task, to which we refer as \emph{master tape}. OpDiLib performs additional treatment for parallel code in \lstinline{f} in a fully automatic fashion and ultimately embeds appropriate reverse handling into the master tape.
In order to be coupled with OpDiLib, an AD tool must additionally
\begin{itemize}
	\item allow for creation and destruction of tapes at runtime,
	\item treat tapes as thread-local objects that can be exchanged at runtime, and on which the classical operations such as recording or evaluation can be performed individually,
	\item provide access to tape positions and support positional evaluation and positional reset of tapes,
	\item have capabilities to embed custom function calls into the tape evaluation, commonly referred to as \emph{external functions} \cite{Kowarz2008}.
\end{itemize}
Furthermore, the AD tool must be thread-safe for use in an OpenMP parallel environment.
\begin{itemize}
	\item Recordings that are performed in parallel on different tapes must be thread-safe. Particularly, the simultaneous distribution of indices to active variables in multiple threads must be safe.
	\item The AD tool must support atomic operations on the adjoint vector. It is sufficient if each evaluation uses atomic access by default. However, to allow for additional performance optimizations, it is preferred that the adjoint access mode can be specified for each individual positional tape evaluation. The most obvious optimization is that serial parts of the code are not evaluated with atomic updates.
\end{itemize}
The full interface that has to be implemented for coupling OpDiLib with an operator overloading AD tool is shown in Listing \ref{listing:ad_tool_interface}. It reflects the AD tool requirements identified above. Note that any specific types used by the AD tool for tapes and tape positions are abstracted away, hence the various \lsttable{void*} pointers in the interface.
\lstset{basicstyle=\footnotesize\ttfamily}
\begin{lstlisting}[caption={Interface that guides the implementation of an AD tool wrapper.}, label={listing:ad_tool_interface}]
struct ADToolWrapperInterface {
	public:
		virtual ~ADToolWrapperInterface() {}

		virtual void init() = 0;
		virtual void finalize() = 0;

		// tape creation and deletion
		virtual void* createTape() = 0;
		virtual void deleteTape(void* tape) = 0;

		// management of thread local tape of caller
		virtual void* getThreadLocalTape() = 0;
		virtual void setThreadLocalTape(void* tape) = 0;

		// position handling
		virtual void* allocPosition() = 0;
		virtual void freePosition(void* position) = 0;
		virtual size_t getPositionSize() = 0;
		virtual std::string positionToString(void* position) = 0;
		virtual void getTapePosition(void* tape, void* position) = 0;

		// tape handling
		virtual bool isActive(void* tape) = 0;
		virtual void setActive(void* tape, bool active) = 0;
		virtual void evaluate(void* tape, void* start, void* end,
				bool useAtomics = true) = 0;
		virtual void reset(void* tape, bool clearAdjoints = true) = 0;
		virtual void reset(void* tape, void* position,
				bool clearAdjoints = true) = 0;
		virtual void pushExternalFunction(void* tape, Handle const* handle) = 0;

		// tape editing
		virtual void erase(void* tape, void* start, void* end) = 0;
		virtual void append(void* dstTape, void* srcTape, void* start,
				void* end) = 0;
};
\end{lstlisting}
\lstset{basicstyle=\small\ttfamily}
In the remaining parts of this section, we discuss some refinements of the AD workflow from Listing \ref{listing:ad_workflow} for advanced use cases.

\paragraph{Multiple recordings and evaluations.} The AD tool may support multiple recordings on the same tape and multiple evaluations of one recorded tape, but OpDiLib does not depend on these features. However, it differentiates them correctly if they are present.

\paragraph{Positional evaluation of the master tape.} If the AD workflow in the serial part of the code involves positional evaluations of the master tape, additional small modifications of the workflow are required in order to get correct results with OpDiLib. As detailed in Section \ref{section:logic}, OpDiLib maintains internal states for the differentiation. State changes are tracked alongside the recording on the classical tapes, and reverted accordingly during evaluation. Hence, if parts of the master tape are evaluated starting from a position that does not match OpDiLib's current state, the state corresponding to that position must be restored first. To that end, OpDiLib's logic layer offers functions for the export and recovery of internal states. These have to be used alongside the export of tape positions and prior to positional evaluation, respectively. The resulting workflow is depicted in Listing \ref{listing:modified_workflow}. 

\lstset{basicstyle=\footnotesize\ttfamily}
\begin{lstlisting}[caption={Modified workflow for positional evaluation of the master tape.}, label={listing:modified_workflow}]
AD::Position startPosition = AD::masterTape->getPosition();

AD::masterTape->startRecording();

// function to be differentiated, involves parallel code
f(inputs, outputs); 

AD::masterTape->endRecording();

AD::Position endPosition = AD::masterTape->getPosition();
OpDiLib::State endState = OpDiLib::logic->exportState();

... // other code

OpDiLib::logic->recoverState(endState);
AD::masterTape->evaluate(endPosition, startPosition);

\end{lstlisting}
\lstset{basicstyle=\small\ttfamily}

\paragraph{Tape activity changes within parallel code.} There are some edge cases that --- unlike we assumed in Listing \ref{listing:ad_workflow} --- require a local interruption of the recording within an otherwise recorded parallel region, for example external functions that are executed on the AD type instead of a scalar type. This is supported by OpDiLib. In general, the tape activity is used by OpDiLib as an indicator whether or not to handle AD events, which has multiple implications. First, recordings may only be performed inside a parallel region whose entry is recorded because otherwise no parallel recording environment is prepared. Second, the tape activity at any begin event should be identical to the tape activity at the matching end event. Third, any synchronization that takes place during interruptions of the recording is not handled by OpDiLib. Tape activity changes are therefore discouraged. An experienced user, on the other hand, can take advantage of this and eliminate synchronization from the tapes that is not required during the reverse pass. To give an example, the following code showcases a barrier that does not result in a barrier in the reverse pass.
\lstset{basicstyle=\footnotesize\ttfamily}
\begin{lstlisting}
AD::threadLocalTape->setPassive();
#pragma omp barrier
AD::threadLocalTape->setActive();
\end{lstlisting}
\lstset{basicstyle=\small\ttfamily}

\paragraph{External functions involving parallelism.} While OpDiLib uses external functions to embed state changes and reverse parallelization into classical tapes, they are originally designed as a tool for users to provide custom derivatives for code, say a function \lstinline{g}, where blackbox differentiation is inaccurate or costly in terms of memory or runtime \cite{Kowarz2008}. A typical example for this is the differentiation through a solver for a linear system of equations \cite{SagebaumGNLL2013}.
Instead of the statement-level recording, user-provided code for the differentiation of \lstinline{g} is embedded into the tape.
Transitioning to the parallel case, the user-provided derivative code does not only replace the derivative computation but also the reverse parallelism that would otherwise be deduced automatically by OpDiLib. This gives experienced users another opportunity to fine-tune the performance.
OpDiLib is not aware whether the tapes it evaluates contain usual statement-level recordings or external functions. Hence, support for external functions involving parallelism is mostly an issue of the underlying AD tool. Particularly, its external function tools must be thread-safe, which involves declaration of inputs and outputs, tape activity changes, the registrations of the external function on the tapes, and where necessary, barrier synchronization between these steps. From OpDiLib's point of view, each thread that registered the external function on its tape in the forward pass results in a thread that participates in the evaluation of the custom derivative code in the reverse pass.

\section{Event-Based Differentiation Logic}
\label{section:logic}

We present the event-based differentiation logic that is implemented in OpDiLib's logic layer. Throughout, we identify requirements for the event generation mechanism.

\subsection{Handling of OpenMP Constructs}
\label{subsec:transformation_of_openmp}

\begin{figure}
\centering
\includegraphics[width=0.8\textwidth]{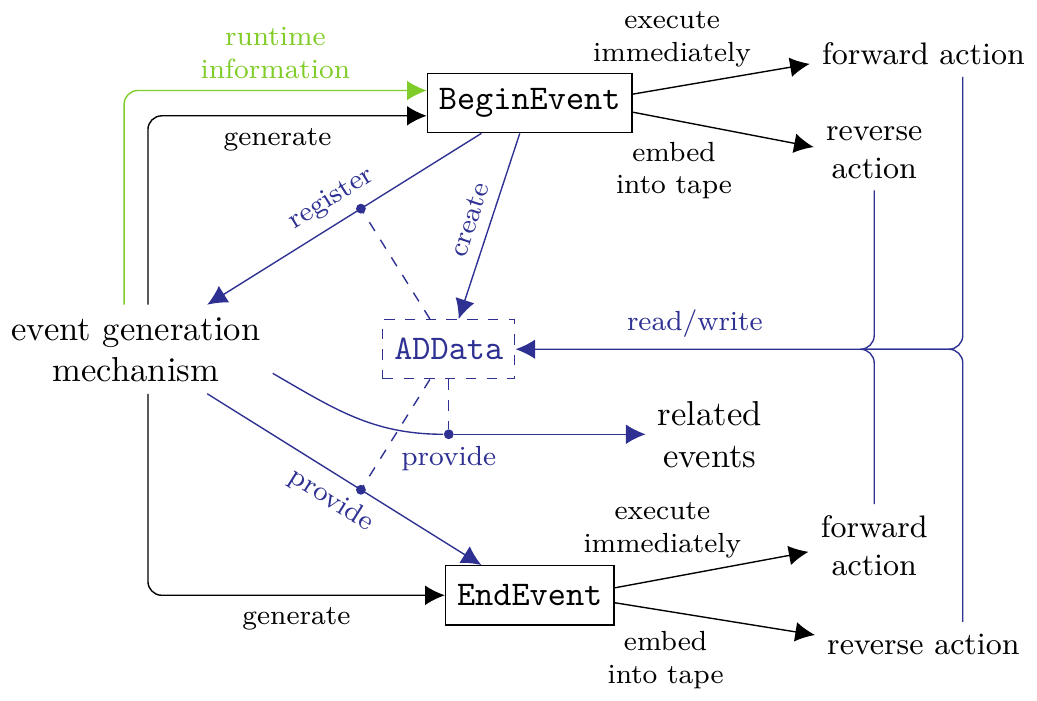}
\caption{Event-based framework for the differentiation of OpenMP. \textbf{Black arrows} indicate the code execution patterns, \textcolor{Blue}{\textbf{blue arrows}} indicate the event-related AD data flow and usage. Additional data flow is indicate by the \textcolor{LimeGreen}{\textbf{green arrow}}.}
\label{figure:event_based_ad_framework}
\end{figure}

We propose to implement the AD handling of OpenMP as reactions to events that occur in pairs and indicate the beginning and end of OpenMP constructs alongside the OpenMP workflow. This is inspired by the execution model events introduced in the OpenMP 5.0 specification \citepalias{OpenMP50}. However, there is a multitude of execution model events, and a lot of internal information is passed to the callbacks that can be registered for these events. Not all of it is relevant for differentiation. For the purpose of AD, we derive from the execution model events a reduced set of events with correspondingly adapted data, to which we refer as \emph{AD events} in the following. AD events admit a joint AD handling of similar OpenMP constructs. Furthermore, they do not depend on OMPT and can be used together with other event generation mechanisms, for example the one presented in Section \ref{section:macro_backend}.
An overview over the event-based framework is provided in Figure \ref{figure:event_based_ad_framework}.
For each AD event, the handling consists of a \emph{forward action} and a \emph{reverse action} that are accompanied by appropriate \emph{AD data}. The forward action contains additional code for the forward pass and the reverse action contains reverse OpenMP logic in the shape of an external function. Upon encountering the event, the forward action is executed immediately and the reverse action is pushed onto the thread's tape.
AD data encompasses any data required for the AD handling in the forward and reverse actions of matching begin and end events. It is usually created in the forward action of a begin event, passed to the matching end event, and provided to the reverse actions of either of both events as external function data. In addition, the AD data is provided to related AD events.
Besides generating the AD events, the event generation mechanism should provide them with runtime information as required for the AD handling and link matching begin and end events as well as families of related events by providing them with the correct AD data.

Differentiation of OpenMP constructs within the framework presented in Figure \ref{figure:event_based_ad_framework} is enabled by the observation that code augmentations for AD are required solely at the beginning and end of OpenMP constructs, together with the respective positions in the reverse pass. In the remaining parts of this section, we show how known operator overloading approaches fit into this framework, how approaches known from source transformation can be transferred to an operator overloading setting, and how these mechanisms can be extended to cover additional OpenMP constructs, up to what is specified in OpenMP 2.5 \citepalias{OpenMP25}. In doing so, we identify also required runtime information.

\begin{figure}[t]
\centering
\includegraphics{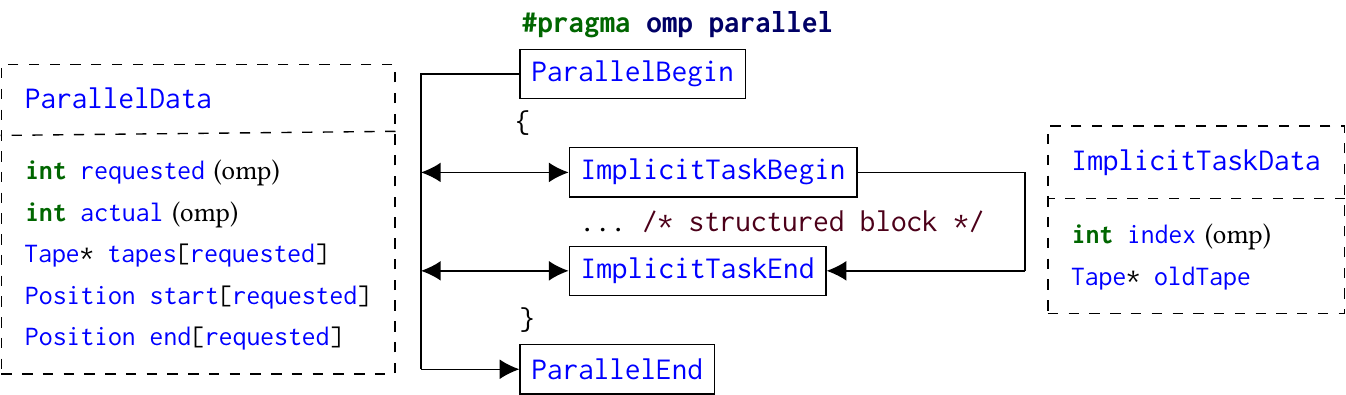}
\caption{Parallel region related AD events in the intuitive ordering with respect to each other and the OpenMP workflow, as well as corresponding AD data flow. Runtime information that the event generation mechanism is required to provide is tagged with (omp).}
\label{figure:parallel_region_events}
\end{figure}

\begin{figure}[t]
\centering
\includegraphics{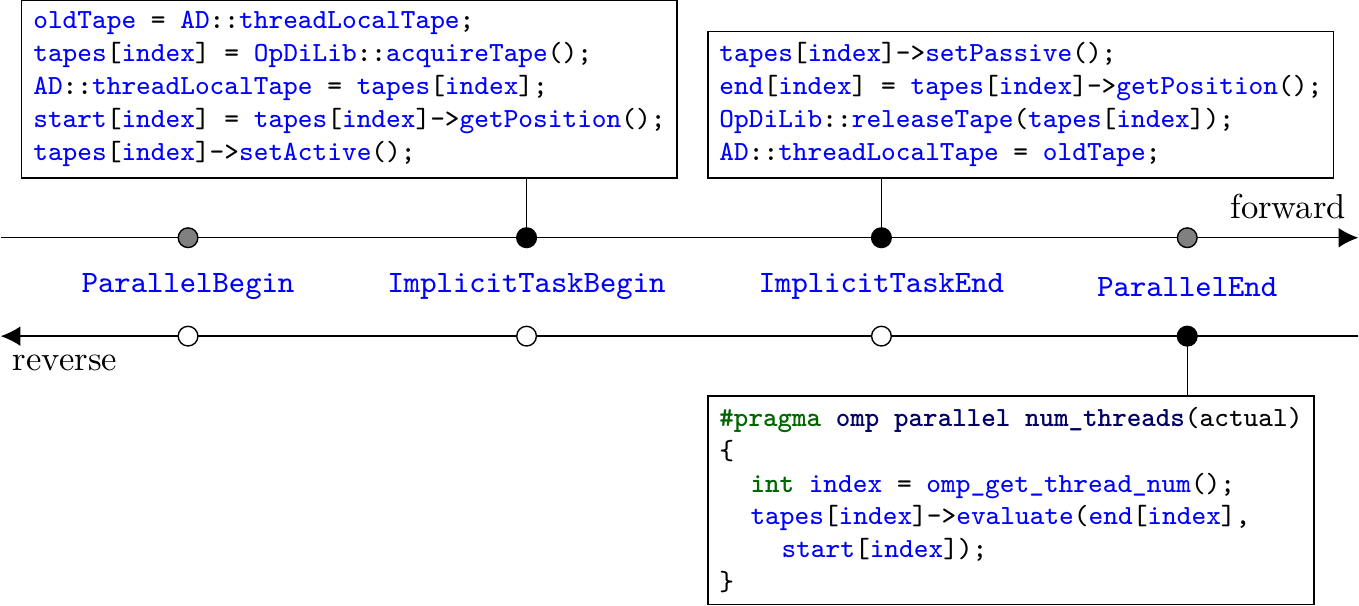}
\caption{Forward and reverse actions of the parallel region AD event family. Actions without description are either empty (white) or contain management tasks like data allocation, data initialization, and the reverse action external function push (grey). The corresponding AD data is displayed in Figure \protect{\ref{figure:parallel_region_events}}.}
\label{figure:parallel_region_actions}
\end{figure}

\begin{figure}
\begin{minipage}{0.4\textwidth}
\includegraphics[width=\textwidth]{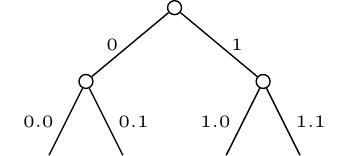}

\vspace{0.2cm}
\end{minipage}
\hfill
\begin{minipage}{0.53\textwidth}
\lstset{basicstyle=\footnotesize\ttfamily}
\begin{lstlisting}
#pragma omp parallel num_threads(2)
{
  workA();
/@\hphantom{ww\,}@/#pragma omp parallel num_threads(2)
  {
    workB();
  }
}
\end{lstlisting}
\lstset{basicstyle=\small\ttfamily}
\end{minipage}
\lstset{basicstyle=\footnotesize\ttfamily}
\caption{Nested parallel regions (right) with the corresponding nesting tree (left). Non-leaf nodes indicate forks and edges correspond to implicit tasks, numbered relative to the parent task. If \lstinline{workA} is imbalanced so that the implicit tasks 0.0 and 0.1 are finished before 1.0 and 1.1 are started, a simple tape management might use the same tapes to record 1.0 and 1.1. If indices are reused, the same index pools would be used for temporaries throughout \lstinline{workB} in the left and right half of the nesting tree. As all four nested tasks are likely evaluated in parallel in the reverse pass, this would create data races.}
\lstset{basicstyle=\small\ttfamily}
\label{figure:hierarchical_taping}
\end{figure}

\paragraph{Parallel regions and implicit tasks.} Appropriate AD handling in an operator overloading setting was established in \cite{Kowarz2008} and implemented in ADOL-C \cite{BischofGKW2008}. Threads perform concurrent recordings on different tapes in the forward pass, and evaluate them in parallel in the reverse pass. We use this as a detailed example for AD handling within the event-based framework illustrated in Figure \ref{figure:event_based_ad_framework}. The corresponding AD event family with associated AD data is displayed in Figure \ref{figure:parallel_region_events}, with the respective forward and reverse actions detailed in Figure \ref{figure:parallel_region_actions}.
The preparation of the parallel recording environment and the parallel tape evaluation are handled by the \lsttable{ParallelBegin} and \lsttable{ParallelEnd} events, respectively, whereas the management of individual tapes and tape position is the responsibility of \lsttable{ImplicitTaskBegin} and \lsttable{ImplicitTaskEnd} events. As an advancement to what is reported on in \cite{Kowarz2008}, and similar to what is achieved in \cite{KalerSXLCPK2021}, the implementation shown in Figure \ref{figure:parallel_region_actions} can be extended to cover also nested parallelism. This is achieved by appropriate tape management. Without nesting, it suffices if the mechanism behind the \mbox{\lsttable{OpDiLib::acquireTape()}} and \lsttable{OpDiLib::releaseTape()} calls ensures that no tape is used for recording by two threads simultaneously. With nesting, we provide tapes according to the position of the implicit task in the nesting tree. Tapes can be reused in subsequent parallel regions encountered by the same parent task but not across nesting levels or parent tasks. The resulting hierarchical arrangement of tapes can be seen as an extension of the nested taping described in \cite{Kowarz2008} and has similarities to the series-parallel tape introduced in \cite{KalerSXLCPK2021}. Figure \ref{figure:hierarchical_taping} visualizes this and highlights a possible issue of tape reuse.

\paragraph{Sync regions.} It is known from source transformation that the reverse counterpart of a sync region, like an explicit \lstinline{#pragma omp barrier} directive, is also a barrier directive, with no additional treatment in the forward pass \cite{Foerster2014}. We introduce a \lstinline{SyncRegionBegin} event and a \lstinline{SyncRegionEnd} event and place the reverse barrier in the reverse action of one of these events.
This covers all types of barriers, including implicit barriers at the end of OpenMP worksharing directives, where we require the event generation mechanism to produce events only if there is no \lstinline{nowait} clause.

\paragraph{Worksharing.} Worksharing directives like \lstinline{#pragma omp for}, \lstinline{#pragma omp single}, and \lstinline{#pragma} \lstinline{omp sections} define independent units of work that are scheduled for execution by specific threads at runtime. Since each thread is equipped with its own tape already, the specific worksharing is reflected in the recordings performed on the different tapes; the same worksharing is automatically achieved in the reverse pass simply by evaluating the tapes in parallel. This means that neither \mbox{\lstinline{#pragma omp section}} declarations of units of work nor \lstinline{schedule} clauses require additional treatment, whereas appropriate schedule reversals have received particular attention in source transformation \cite{HueckelheimH2021}. Additional augmentations for worksharing directives in \cite{Foerster2014} are either specific to source transformation or part of a joint reversal approach, where the memory footprint is reduced by checkpointing techniques as part of the transformation. As is common in operator overloading tools, a user has to employ such techniques manually. \lsttable{WorkBegin} and \lsttable{WorkEnd} events are reserved for future augmentations of worksharing directives.

\begin{figure}
\centering
\includegraphics[width=\textwidth]{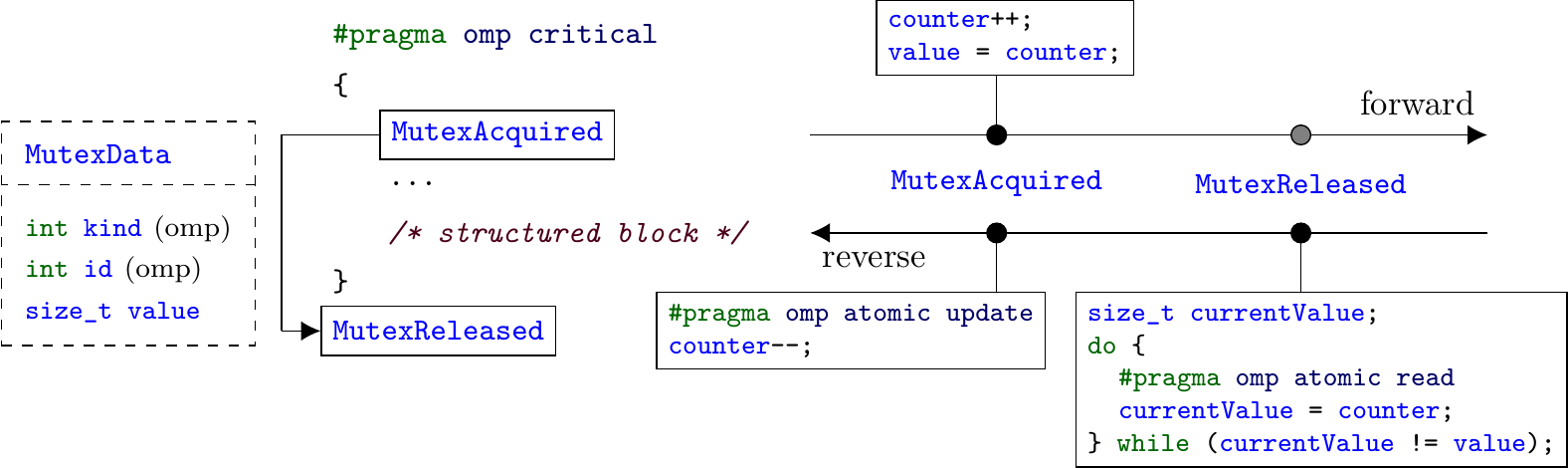}
\lstset{basicstyle=\footnotesize\ttfamily}
\caption{Mutex related AD events in the intuitive ordering with respect the OpenMP workflow, as well as corresponding AD data flow (left) and corresponding forward and reverse actions (right). \lsttable{counter} references a global integer that depends on \lsttable{kind} and \lsttable{id}. The omitted action (grey) contains only the external function push.  The increment of the counter is part of the \lsttable{MutexAcquired} forward action, which we require --- unlike the \lsttable{MutexReleased} forward action --- to be protected by the underlying mutex.}
\lstset{basicstyle=\small\ttfamily}
\label{figure:mutex_events_actions}
\end{figure}

\paragraph{Mutexes.} To differentiate mutex-type synchronization correctly, we have to reverse the specific execution order produced in the forward pass. This is reflected in a counter-based AD approach developed for critical regions in source transformation \cite{Foerster2014}, where entering the critical region corresponds to acquiring the mutex and leaving the critical region corresponds to releasing it. Figure \ref{figure:mutex_events_actions} shows how it can be implemented with the help of \lsttable{MutexAcquired} and \lsttable{MutexReleased} AD events in our framework.
We make the observation that this concept can be used to differentiate any synchronization of mutex type and apply it for critical regions, locks, nested locks, ordered clauses and reduction clauses. Counters are associated with mutexes via the mutex \lsttable{kind} and a unique mutex \lsttable{id} that have to be provided by the event generation mechanism. The current values of all counters are part of OpDiLib's state that we mentioned before in Section \ref{section:ad_tool} in connection with Listing \ref{listing:modified_workflow}.

\paragraph{Reductions.} Since the OpenMP 4.0 specification \citepalias{OpenMP40}, user-defined reductions can be declared, which can be leveraged for reductions involving data types that are used in operator overloading AD tools. However, a declaration of a custom reduction alone is not sufficient to obtain correct reverse mode derivatives.
Reductions are performed in OpenMP in two stages \citepalias{OpenMP52}. First, each thread that participates in the reduction accumulates its results in a private variable. At the end of the OpenMP construct with the reduction clause, the threads' individual results are then accumulated into the shared reduction target, usually involving several threads and shared intermediate variables, for example in a tree-like reduction. From an AD point of view, the operations in the course of the reduction are recorded just as any other operations on the tapes of the threads that execute them. However, access to shared intermediate variables and the final reduction target requires synchronization, which must be turned into reverse synchronization again. We require that a thread receives a \lsttable{MutexAcquired} event before contributing to a shared variable, and a \lsttable{MutexReleased} event afterwards; just as if access to the shared variable was safeguarded by a mutex. To ensure that no thread begins with the reversal of the structured block of the OpenMP construct before the reversal of the shared reduction stage has progressed to a point where all contributions have been propagated to the private reduction variable of that thread, we require that additional \lsttable{SyncRegionBegin} and \lsttable{SyncRegionEnd} events are emitted between the two reduction phases. This is important for threads that did not participate in the shared reduction phase and did not receive mutex AD events.

\paragraph{Further clauses and directives that do not require treatment.} The \lstinline{#pragma omp master} directive does not involve any synchronization; it only declares that the code will be executed by a specific thread, and hence, recorded on a specific tape. Thus, it will also be evaluated by only one thread. This is in line with the results on source transformation from \cite{Foerster2014}. OpenMP allows combined constructs, usually a parallel directive combined with a worksharing construct, for example \lstinline{#pragma omp parallel for}. Such combined constructs are supported automatically by the logic presented above as long as all involved mechanisms generate their corresponding AD events. Furthermore, we do not have to add special treatment for data-sharing attribute clauses such as \lstinline{private} or \lstinline{firstprivate} or data copying clauses such as \lstinline{copyprivate}. The resulting copy operations are recorded in the same fashion as any other operations. Source transformation tools, on the other hand, have to implement treatment for such clauses \cite{Foerster2014,HueckelheimHSM2019,HueckelheimH2021}.

\paragraph{Atomics.} The OpenMP specification allows \lstinline{#pragma omp atomic} directives only for scalar types, which excludes the custom types that are used for active variables in an operator overloading code. Many of the remaining use cases for atomics, such as computations on non-AD floating point types or increments of counters, include synchronization that does not impact on the correctness of the derivative. Therefore, we do not add AD handling of atomic directives. In theory, synchronization due to atomics could also be handled by mutex AD events; in fact, atomics generate \textit{mutex-acquired} and \textit{mutex-release} execution model events. In practice, patterns like busy-waits would quickly blow up the mutex counters and are more efficiently handled by custom mutex events that reflect the synchronization mechanism. If a pattern such as the busy-wait synchronization in Figure \ref{figure:mutex_events_actions} was used in the forward pass, one \lsttable{MutexAcquired} event could be emitted upon a completed busy-wait, with the corresponding \lsttable{MutexReleased} event placed after decrementing the counter. Note that missing AD handling of atomics does not prevent us from using them in the reverse pass for atomic updates on adjoint variables or for synchronization as displayed in Figure \ref{figure:mutex_events_actions}.

\paragraph{Flush directive.} The \lstinline{#pragma omp flush} directive enforces a consistent view of the threads on the memory, either in general or with respect to specific variables. As can be seen e.\,g.~in the examples in the OpenMP 2.5 specification \citepalias{OpenMP25}, it can be used to implement quite irregular synchronization patterns. We see no way to handle the flush directive without implicitly turning it into a stronger kind of synchronization like atomics, mutexes, or barriers, effectively eliminating its advantages. Regarding the reverse pass, no issues with inconsistent views of memory should occur if atomic updates are used on adjoint variables. Otherwise, we offer a reverse flush that the user may push to the tapes manually, in the same spirit as the reverse barrier described in Section \ref{subsec:data_access_patterns}.

\paragraph{Higher order derivatives.} In the adjoints of tangents  approach, the forward mode of AD is used for all derivative orders except the outermost one. This strategy is not only favoured for its efficiency properties \cite{GriewankW2008} but also supported by the differentiation logic presented above. The adjoints of adjoints approach, on the other hand, applies the reverse mode to the reverse mode, which involves recording of an evaluation that --- in a parallel context --- likely involves atomic updates on adjoint variables. This conflicts with missing AD handling of atomics. It also gives rise to further implementational issues in the event generation mechanism that are discussed in Section \ref{section:backend}. Therefore, adjoints of adjoints are not supported by OpDiLib.

\subsection{Shared Reading and Atomic Adjoints}
\label{subsec:data_access_patterns}

OpDiLib itself cannot automatically identify whether or not atomic updates of adjoint variables are required. In particular, unlike source transformation tools \cite{HueckelheimH2021}, it cannot distinguish between \lstinline{private} and \lstinline{shared} variables specified on OpenMP constructs via the corresponding data-sharing attribute clauses. The reason is that there are no execution model events associated with these clauses, and hence, there is no solution for this issue in the scope of this work. In order to make OpDiLib as applicable as possible, we do not want to impose any restrictions on data access patterns and we do not want to require users to restructure their codes to get a first working version. Therefore, we default to atomic updates on adjoint variables. On the other hand, a user should have as much flexibility as possible to disable atomic adjoints manually where they are not needed. To that end, OpDiLib tracks an adjoint access mode per thread alongside the recording as part of the parallel region's associated AD data, which the event generation mechanism is required to make availabe. The adjoint access options are \lsttable{Atomic}, which is the default for OpDiLib, or \lsttable{Classical}, which is the non-atomic counterpart. The user may control the adjoint access mode by calls to the \lstinline{setAdjointAccessMode} routine of the logic layer at arbitrary points in the code.
The recording that is performed after the call until the next change of the adjoint access mode in that thread is then evaluated in the reverse pass according to the chosen mode.
This can be used to disable atomics globally, for example if the reversal strategy from \cite{KalerSXLCPK2021} is implemented in the underlying AD tool. It can also be applied to mark individual parallel regions as safe for evaluation without atomics, for example if the user decides to make thread-private copies of all inputs to the parallel region as in \cite{Kowarz2008,BischofGKW2008}, or if a parallel loop has self-adjoint memory access \cite{HueckelheimHSM2019}. In addition, one may isolate any parts of large parallel regions as safe for evaluation without atomics in a fine-grained manner, even across the boundaries of parallel regions. The following example demonstrates that additional synchronization might be required during the reverse pass.
\lstset{basicstyle=\footnotesize\ttfamily}
\begin{lstlisting}
ADType shared_data[2] = ...;
ADType result[2] = ...;
#pragma omp parallel num_threads(2)
{
	int i = omp_get_thread_num();
	OpDiLib::logic->setAdjointAccessMode(OpDiLib::Classical);

	// part without shared reading
	result[i] *= shared_data[i];

	OpDiLib::logic->addReverseBarrier();	
	OpDiLib::logic->setAdjointAccessMode(OpDiLib::Atomic);

	// part with shared reading
	result[i] += shared_data[0] * shared_data[1];
}
\end{lstlisting}
\lstset{basicstyle=\small\ttfamily}
The change of the adjoint access mode in line 12 separates adjacent parts of the code with and without shared reading. During the forward pass, shared reading is not a problem and no synchronization is required. During the reverse pass, however, one thread might finish first with the atomic evaluation of the recording for line 15 and begin with unsafe updates on the adjoint of \lstinline{shared_data[i]} while the other thread is still busy with the safe evaluation of its recording for line 15, resulting in simultaneous safe and unsafe writes to the same variable. This is prevented by introducing a barrier exclusively in the reverse pass by the additional call in line 11. It pushes an external function containing a \lstinline{#pragma omp barrier} statement to the tape, just as in the handling of the \lsttable{SyncRegion} events in Section \ref{subsec:transformation_of_openmp}.

\section{Event Generation Mechanisms}
\label{section:backend}

The AD events introduced in Section \ref{subsec:transformation_of_openmp} have to be generated alongside the OpenMP workflow. In principle, the required augmentation of the source code by AD events could be performed by hand. The objective of OpDiLib's backend layer is to automatize this tedious task, up to the point where no changes to the source code are required. This is achieved with the OMPT backend that is presented in Section \ref{section:ompt_backend}. It makes use of modern OpenMP features around OMPT that allow for an observation and augmentation of the OpenMP workflow. This way, we solve the key challenge of augmenting \lstinline{#pragma} directives without manual annotations of the source code.

As a fallback for compilers without OMPT support, we provide a second backend that hides the augmentations of the source code inside replacement macros for OpenMP \lstinline{#pragma} directives and clauses as well as replacements for OpenMP's runtime functions and additional end macros. They form an alternative interface to OpenMP functionality which we present in Section \ref{section:macro_backend}. It provides access to the same set of AD features as the OMPT backend. Clearly, the macro backend requires the rewriting of all OpenMP directives and runtime function calls. In principle, given the structural similarities of the original constructs and their replacements, this task could be performed by a source transformation preprocessing script in an automatic fashion. Code that is written with the replacements is also fully compatible with the OMPT backend. This means that no additional rewriting is required for a transition to the modern features. As long as the OMPT backend is used, original OpenMP constructs and their respective replacements can be used in the same code.

Full code examples that showcase the use of the two backends can be found in the OpDiLib repository\footnote{\url{https://github.com/SciCompKL/OpDiLib}}.

\subsection{OMPT Backend}
\label{section:ompt_backend}

The OpenMP specification \citepalias{OpenMP52} formalizes the OpenMP workflow by defining \emph{execution model events}. For each event, a custom callback function can be registered that is invoked by the OpenMP runtime at each occurrence of the corresponding event in the respective thread. The callback function receives detailed information about the OpenMP mechanisms and objects involved. In many cases, it also offers possibilities to associate custom data with events. This allows developers to build their own tools on top of OpenMP and integrate them seamlessly into the workflow of the OpenMP runtime. We use these features for the generation of AD events without any modifications of the OpenMP parallel source code. Since the AD events introduced in Section \ref{subsec:transformation_of_openmp} are derived from the execution model events, a key part of a callback implementation is always the transformation to the corresponding AD event. Additionally, OMPT is powerful enough to fulfill further requirements identified throughout Section \ref{section:logic}, for example AD data transport, mutex identifier generation or general access to parallel AD data.

We illustrate how a backend can be implemented with functionality provided by OMPT using the example of a parallel region and its implicit tasks. The OpenMP specification \citepalias{OpenMP52} defines the following order of events for a parallel region.
\begin{itemize}
\item A thread that encounters a parallel construct receives a \emph{parallel-begin} event before any implicit task associated with the parallel region is created.
\item Guided by the requested degree of parallelism, implicit tasks are generated, all of which execute the structured block of the parallel construct. Each implicit task is executed by a dedicated thread. The thread that encountered the parallel region executes the implicit task with index zero.
\item Each thread that participates in the parallel region receives an \emph{implicit-task-begin} event before it starts executing the structured block. After it has finished executing the structured block and has received events associated with the implicit barrier at the end of the parallel region, it receives an \emph{implicit-task-end} event.
\item After the thread that encountered the parallel region has received its \emph{implicit-task-end} event, it receives a \emph{parallel-end} event. After that, it resumes execution of the task in which it encountered the parallel region.
\end{itemize}
This order is inherited by the corresponding AD events. We remark that no ordering is specified for the \emph{parallel-end} event and \emph{implicit-task-end} events of implicit tasks other than the one with index zero. The extreme case is that they occur no sooner than immediately prior to the next \emph{implicit-task-begin} in that thread, in particular after the next \emph{parallel-begin} event. We observed this in the LLVM OpenMP runtime. The implementation of the differentiation logic has to be correct also for these extreme cases.
Listing \ref{listing:parallel_callbacks} displays implementations of the callbacks for \emph{parallel-begin} and \emph{parallel-end} events that match the \lstinline{ompt_callback_parallel_begin_t} and \lstinline{ompt_callback_parallel_end_t} signatures as defined by the OpenMP specification \citepalias{OpenMP52}.
\lstset{basicstyle=\footnotesize\ttfamily}
\begin{lstlisting}[caption={Callbacks for \emph{parallel-begin} and \emph{parallel-end} events.}, label={listing:parallel_callbacks}]
void on_parallel_begin(ompt_data_t*, const ompt_frame_t*,
		ompt_data_t* parallelData, unsigned int requested, int, void const*) {
	parallelData->ptr = OpDiLib::logic->onParallelBegin(requested);
}

void on_parallel_end(ompt_data_t* parallelData, ompt_data_t*, int,
		void const*) {
	OpDiLib::logic->onParallelEnd(parallelData->ptr);
}
\end{lstlisting}
\lstset{basicstyle=\small\ttfamily}
We omitted the names of arguments that we do not use. The callback for the \emph{parallel-begin} event receives information about the \lstinline{requested} parallelism, that is, the number of threads that was specified with the \lstinline{num_threads} clause if present, or the default number of threads instead, as required in Section \ref{subsec:transformation_of_openmp}. Furthermore, the OpenMP runtime allocates a parallel data object of structured type \lstinline{ompt_data_t} that is associated with the parallel region. It has a member \lstinline{void* ptr} that can be used to add custom data. We use it to store the parallel AD data object returned by the logic layer. The callback for the corresponding \emph{parallel-end} event receives the same OMPT data object with the \lstinline{ptr} member as it was assigned in the \emph{parallel-begin} callback. Hence, we can use it to transport the AD data between the matching begin and end events as displayed in Figures \ref{figure:opdilib_architecture} and \ref{figure:event_based_ad_framework} and required in Section \ref{subsec:transformation_of_openmp}. OMPT's data object associated with the current parallel region can also be queried at arbitrary points in the code via the \lstinline{ompt_get_parallel_info} runtime entry point. Thus, we implement access to the parallel AD data object of the current parallel region as is needed in Section \ref{subsec:data_access_patterns} by querying OMPT's parallel data object, and extracting the parallel AD data object from it in the same fashion as in Listing \ref{listing:parallel_callbacks}. Listing \ref{listing:implicit_task_callbacks} displays the implementation of the callback for \emph{implicit-task-begin} and \emph{implicit-task-end} events. It has the \lstinline{ompt_callback_implicit_task_t} signature as required by the OpenMP specification \citepalias{OpenMP52}.
\lstset{basicstyle=\footnotesize\ttfamily}
\begin{lstlisting}[caption={Callback for \emph{implicit-task-begin} and \emph{implicit-task-end} events.}, label={listing:implicit_task_callbacks}]
void on_implicit_task(ompt_scope_endpoint_t endpoint, ompt_data_t* parallelData,
    ompt_data_t* taskData, unsigned int actual, unsigned int index, int flags) {

	if (flags & ompt_task_initial) {
		return;
	}

	if (ompt_scope_begin == endpoint) {
		taskData->ptr = OpDiLib::logic->onImplicitTaskBegin(
			actual, index, parallelData->ptr);
	}
	else {
		OpDiLib::logic->onImplicitTaskEnd(taskData->ptr);
	}
}
\end{lstlisting}
\lstset{basicstyle=\small\ttfamily}
It filters out the special case of the initial implicit task, that is, the task that executes the serial parts of the program. It does not require additional AD treatment. The callback receives all parameters from the OpenMP runtime that are needed for the implicit task AD events, that is, the \lsttable{actual} degree of parallelism (as opposed to the requested one), the \lsttable{index} in the team of threads, and the parallel AD data object. The task data object provided by OMPT is again used to exchange AD data with the corresponding end event.

Table \ref{table:callbacks_to_ad_events} provides an overview over all callbacks that we use to generate the AD events with OpDiLib's OMPT backend. 
\begin{table}
\centering
\caption{Mapping between OMPT callbacks and AD events.}
\label{table:callbacks_to_ad_events}
\begin{tabular}{ll}
\toprule
OMPT callback type & AD events \\
\midrule
\lsttable{ompt_callback_parallel_begin} & \lsttable{ParallelBegin}\\
\lsttable{ompt_callback_parallel_end} & \lsttable{ParallelEnd}\\
\lsttable{ompt_callback_implicit_task} & \lsttable{ImplicitTaskBegin}, \lsttable{ImplicitTaskEnd}\\
\lsttable{ompt_callback_mutex_acquired} & \lsttable{MutexAcquired}\\
\lsttable{ompt_callback_mutex_released} & \lsttable{MutexReleased}\\
\lsttable{ompt_callback_sync_region} & \lsttable{SyncRegionBegin}, \lsttable{SyncRegionEnd}\\
\lsttable{ompt_callback_work} & \lsttable{WorkBegin}, \lsttable{WorkEnd}\\
\lsttable{ompt_callback_reduction} & \lsttable{MutexAcquired}, \lsttable{MutexReleased}\\
\bottomrule
\end{tabular}
\end{table}
Not each execution model event has a dedicated callback function. In contrary, a single callback often handles a large number of events. For example, the \emph{critical-acquired} event due to a \lstinline{#pragma omp critical} directive and the \emph{lock-acquired} event caused by the runtime function \lstinline{omp_set_lock} both result in invocations of the \lstinline{ompt_callback_mutex_acquired} callback. Likewise, an \lstinline{ompt_callback_sync_region} callback handles (amongst others) explicit barrier directives and implicit barriers at the end of, e.\,g., worksharing directives that do not specify a \lstinline{nowait} clause. It is beneficial for AD that similar types of events are collected in one callback for two reasons. First, this often coincides with the fact that all execution model events collected by the same callback also require the same treatment from an AD point of view and can be forwarded to the same module of the logic layer. Second, we do not have to distinguish whether the execution model event is caused by a \lstinline{#pragma} directive, the presence or absence of clauses, a runtime function call, or other mechanisms. This allows us to conveniently establish AD support for a large range of OpenMP features. 

In the view of further requirements identified in Section \ref{subsec:transformation_of_openmp}, the OpenMP runtime passes unique mutex identifiers to mutex callbacks, which we can immediately reuse to identify and track mutexes in the logic layer. Pending \emph{mutex-released} events in one thread  might occur only after \emph{mutex-acquired} events for the same mutex in another thread. Unlike the handling of the \emph{mutex-acquired} event, the handling of the \emph{mutex-released} event is not protected by the corresponding mutex, which is regarded in the implementation of the differentiation logic. The reduction callback is special among the cases in Table \ref{table:callbacks_to_ad_events} in the sense that it produces mutex AD events. According to the OpenMP specification \citepalias{OpenMP52}, the \textit{reduction-begin} event is received before the task begins to perform load and stores on shared reduction variables, and the \textit{reduction-end} event is received after these are completed. We use these to generate the mutex AD events required in Section \ref{subsec:transformation_of_openmp}. Thus we record the reduction operations as if they were performed in an equivalent serial manner, that is, we choose a total order for all reduction events that preserves the partial order indicated by the reduction tree. Reversal is then performed according to this total order.
Conveniently, the LLVM OpenMP runtime produces sync region events for a so-called implementation specific barrier that separates the execution of the structured block and the accumulation in private reduction variables from the part of the reduction that actually involves synchronization. These events are detected and handled by OpDiLib like any other sync region. Hence, when using this OpenMP runtime, it is ensured that no thread begins with the reversal of the structured block before all reduction operations are reverted. This is important for threads that did not participate in the reduction and did not receive mutex events. This implementation specific barrier is not guaranteed by the OpenMP specification. Besides its usefulness for AD, it provides insights in the reduction process that should also be of interest to performance monitoring tools. Therefore, we think that it is a desirable feature in general and would like to see it in future versions of the OpenMP specification. If OpDiLib is used together with other OMPT implementations, it might be necessary to add a reverse barrier by hand, or to use semi-automatic techniques similar to those in the macro backend.
We close this section with a discussion of additional issues due to the use of OMPT. 

\paragraph{Execution model events due to internal synchronization.} OpDiLib uses OpenMP's lock facilities for internal synchronization purposes. From the OpenMP runtime's point of view, these are not any different from locks in user code and hence produce the same execution model events and cause invocations of the same callbacks. Besides the unnecessary performance cost of differentiating the logic layer's internals, there is also a more imminent issue. If the user successfully sets a lock, an \lstinline{ompt_callback_mutex_acquired} callback is dispatched and generates a \lsttable{MutexAcquired} AD event. However, the handling of this event itself requires the acquisition of an internal lock that invokes the same callback, resulting in an infinite recursion. 
To fix this, the logic layer keeps track of inactive mutexes. These are then filtered out in the AD event handling. An experienced user can also use these mechanisms to mark mutexes from user code as inactive by a call \lstinline{OpDiLib::logic->registerInactiveMutex(mutexKind, mutexId)}.
These are then disregarded for the differentiation. This is also recommended for mutexes introduced in the underlying AD tool for the purpose of thread-safety.

\paragraph{Execution model events during reversal.} Since OpDiLib embeds OpenMP directives into the tape, execution model events are also produced during the reverse pass. However, the tapes are not in recording mode at this point. This is detected in the logic layer and the AD handling is skipped.
We remark --- despite the limited practical relevance --- that execution model events during reversal might have an application for higher order derivatives that are computed in an adjoints of adjoints fashion, see the previous discussion at the end of Section \ref{subsec:transformation_of_openmp}. In an application of OpDiLib to itself, the execution model events during an evaluation for the higher order type could be used to perform the recording for the lower order type. This relates also to the next topic of discussion.

\paragraph{Multiple OMPT tools.} Suppose we want to apply OpDiLib to an OpenMP parallelized code that uses an OMPT tool already, for example for performance monitoring purposes.
The simultaneous use of multiple OMPT tools is beyond the scope of the OpenMP specification \citepalias{OpenMP52}. To overcome this limitation, there was work in the direction of multiplexing of OMPT tools \cite{ProtzeCCM2019} that is now part of \verb|clang|. We will conduct further tests to see whether it can be used for our purposes. Otherwise, the macro backend may serve as a fallback solution.

\subsection{Macro Backend}
\label{section:macro_backend}

Without OMPT, we are not aware of any way to change the behaviour of the original OpenMP constructs in the scope of the program. Instead, we provide an alternative interface to OpenMP functionality. Code that is written according to this interface can then also be differentiated if a compiler without OMPT support is used. Specifically, we introduce macros that have to be used instead of the usual OpenMP directives and clauses, as well as functions that have to be used instead of the usual OpenMP runtime functions. This provides a semi-automatic way of differentiating OpenMP parallel code with operator overloading. The replacements always contain the original OpenMP construct but embed it into additional code that generates the required AD events, and the replacements follow mostly the same structure as the original OpenMP constructs.

\paragraph{Runtime functions.} For each runtime function \lstinline{omp_*}, we introduce a replacement \lstinline{opdi_*}. For example, \lstinline{omp_set_lock} is replaced by \lstinline{opdi_set_lock}. The replacement has the same signature as the original function. Its implementation always forwards the call to the original runtime function, but surrounds it with appropriate generation of AD events, as showcased in Listing \ref{listing:set_lock_replacement}.
\lstset{basicstyle=\footnotesize\ttfamily}
\begin{lstlisting}[caption={\lstinline{omp_set_lock} runtime function replacement.}, label={listing:set_lock_replacement}]
void opdi_set_lock(omp_lock_t* lock) {
	omp_set_lock(lock);
	OpDiLib::logic->onMutexAcquired(OpDiLib::Lock,
			OpDiLib::backend->getLockIdentifier(lock));
}
\end{lstlisting}
\lstset{basicstyle=\small\ttfamily}
As can be seen in Listing \ref{listing:set_lock_replacement}, the generation of mutex identifiers, that was previously performed by the OMPT runtime, now requires a dedicated custom implementation in the backend layer.
For reasons of consistency, we offer replacements also for runtime functions that do not require AD treatment. A user code that should compile both with and without OpDiLib can then use a global macro to decide whether the \lstinline{omp_} or \lstinline{opdi_} prefix is used for the runtime function calls, without keeping track of the necessity of this switch for each individual runtime function.

\paragraph{Directives.} We replace each OpenMP directive \lstinline{#pragma omp directive ...} by a custom macro \mbox{\lstinline{OPDI_DIRECTIVE(...)}}. The idea is that the same clauses and hints that were passed to the original directive can also be passed to the macro, possibly again in the shape of their replacements. The macro expands to the orignal OpenMP directive surrounded by corresponding AD treatment. This is showcased for the barrier directive in Listing \ref{listing:barrier_replacement}.
\lstset{basicstyle=\footnotesize\ttfamily}
\begin{lstlisting}[caption={Barrier replacement macro.}, label={listing:barrier_replacement}]
#define OPDI_BARRIER(...) \
	OpDiLib::logic->onSyncRegionBegin(OpDiLib::Barrier); \
	_Pragma(omp barrier __VA_ARGS__) \
	OpDiLib::logic->onSyncRegionEnd(OpDiLib::Barrier);
\end{lstlisting}
\lstset{basicstyle=\small\ttfamily}
The situation is more involved for OpenMP directives that are followed by a structured block, for example \lsttable{#pragma omp for}. This has two reasons. First, it might be necessary to generate AD events within the structured block. Second, it might be necessary to spawn AD events after the structured block. We make heavy use of a privatization technique that was already applied to address these issues for the handling of parallel regions in ADOL-C \cite{Kowarz2008,BischofGKW2008}. Augmentations at the beginning and end of the structured block can be generated as side effects of constructors and destructors of objects that are privatized via data-sharing clauses. These privatizations can be generated as part of both replacement macros and replacement clauses. However, this technique is not applicable or practical in all situations. First, it covers only OpenMP directives that support data-sharing clauses. Furthermore, the OpenMP specification leaves the respective order of constructor and destructor calls unspecified, does not support multiple privatizations of the same variable on one directive, and restricts subsequent privatization in nested directives to some extent \citepalias{OpenMP52}. Privatization is the only approach we are aware of to design replacement clauses with side effects on the recording environment. Thus, their limitations give rise to some special cases which we discuss below. To find a remedy for directives, we complement each \lsttable{OPDI_DIRECTIVE(...)} replacement that is followed by a structured block by a corresponding \lsttable{OPDI_END_DIRECTIVE} macro that has to be placed after the structured block. This is showcased in Listing \ref{listing:end_macro} and resembles the Fortran way of using OpenMP directives rather than the \cpp one.

\lstset{basicstyle=\footnotesize\ttfamily}
\begin{minipage}{0.53\textwidth}
\begin{lstlisting}
#pragma omp parallel num_threads(4)
{
/@\hphantom{ww\,}@/#pragma omp for schedule(guided, 1)
	for (int i = 0; i < N; ++i) {
		...
	}

}
	
\end{lstlisting}
\end{minipage}
\hfill
\begin{minipage}{0.45\textwidth}
\begin{lstlisting}
OPDI_PARALLEL(num_threads(4))
{
	OPDI_FOR(schedule(guided, 1))
	for (int i = 0; i < N; ++i) {
		...
	}
	OPDI_END_FOR
}
OPDI_END_PARALLEL
\end{lstlisting}
\end{minipage}
\vspace{-1em}
\begin{lstlisting}[caption={Example code on the left and equivalent code with replacement and end macros on the right.}, label={listing:end_macro}]
\end{lstlisting}
\lstset{basicstyle=\small\ttfamily}
End macros also provide important degrees of flexibility in our implementation with respect to future adaptions. Table \ref{table:replacement_directive_ad} provides an overview over all replacement directives and the corresponding end macros.
\begin{table}
\centering
\caption{Replacement macros and corresponding end macros.}
\label{table:replacement_directive_ad}
\begin{tabular}{lll}
\toprule
OpenMP directive & replacement macro & end macro \\
\midrule
\lsttable{#pragma omp barrier ...} & \lsttable{OPDI_BARRIER(...)} & \\
\lsttable{#pragma omp parallel ...} & \lsttable{OPDI_PARALLEL(...)} & \lsttable{OPDI_END_PARALLEL} \\
\lsttable{#pragma omp for ...} & \lsttable{OPDI_FOR(...)} & \lsttable{OPDI_END_FOR} \\
\lsttable{#pragma omp sections ...} & \lsttable{OPDI_SECTIONS(...)} & \lsttable{OPDI_END_SECTIONS} \\
\lsttable{#pragma omp single ...} & \lsttable{OPDI_SINGLE(...)} & \lsttable{OPDI_END_SINGLE} \\
\lsttable{#pragma omp single nowait ...} & \lsttable{OPDI_SINGLE_NOWAIT(...)} & \lsttable{OPDI_END_SINGLE} \\
\lsttable{#pragma omp critical ...} & \lsttable{OPDI_CRITICAL(...)} & \lsttable{OPDI_END_CRITICAL} \\
\lsttable{#pragma omp critical (name)} \lsttable{...} & \lsttable{OPDI_CRITICAL_NAME(name, ...)} & \lsttable{OPDI_END_CRITICAL} \\
\lsttable{#pragma omp ordered ...} & \lsttable{OPDI_ORDERED(...)} & \lsttable{OPDI_END_ORDERED} \\
\lsttable{#pragma omp master ...} & \lsttable{OPDI_MASTER(...)} & \lsttable{OPDI_END_MASTER} \\
\lsttable{#pragma omp section ...} & \lsttable{OPDI_SECTION(...)} & \lsttable{OPDI_END_SECTION} \\
\bottomrule
\end{tabular}
\end{table}
Among the replacement directives there are two special cases. The first are named critical regions. The \lsttable{critical} directive does not support data-sharing clauses. However, the supplied name is essential for the generation of a unique mutex identifier so that is has to be provided via a dedicated macro. Second, while the \lsttable{single} directive supports data-sharing clauses, these apply only to the single thread that executes the structured block, which is not sufficient for the treatment of the implicit barrier that regards all threads. Hence, a dedicated macro for \lsttable{single} directives with a \lsttable{nowait} clause is introduced. Note that the special replacements still use the same respective end macro.

\paragraph{Clauses.} Most clauses and hints do not require replacements and are passed to the replacement directives in their original form. Besides the special replacement directives discussed above, we provide a \lsttable{OPDI_NOWAIT} replacement clause that has to be used instead of \lsttable{nowait} for correct treatment of the corresponding implicit barrier. Furthermore, any directive that specifies \lsttable{reduction(...)} clauses has to specify the \lsttable{OPDI_REDUCTION} macro once alongside the otherwise unmodified reduction clauses. It provides the reverse barrier that separates the threadprivate reduction phase from the shared, tree-like reduction phase. We remark that this macro has to be provided as a supplement instead of a replacement because multiple reduction clauses on one directive are allowed but multiple privatizations of the same variable on one directive are forbidden \citepalias{OpenMP52}.

\paragraph{Reductions.} While the \lsttable{OPDI_REDUCTION} macro handles the required reduction barrier, it does not handle the mapping of reductions to mutex AD events. In the macro backend, we associate reductions with actual nested locks so that the required mutex AD events are produced by setting and unsetting these locks. This logic is hidden inside a replacement macro for the declaration of custom reductions. Since any reduction involving AD types is of custom type, this covers all the relevant cases. The replacement macro reads \lstinline{OPDI_DECLARE_REDUCTION(OP_NAME, TYPE, OP, INIT)} and expands to code as displayed in Listing \ref{listing:declare_reduction}.
\lstset{basicstyle=\footnotesize\ttfamily}
\begin{lstlisting}[caption={Schematic reduction declaration.}, label={listing:declare_reduction}]
#pragma omp declare \
	reduction(OP_NAME : TYPE : \
		OpDiLib::Reducer(omp_out) \
			= OpDiLib::Reducer(omp_out) OP OpDiLib::Reducer(omp_in)) \
	initializer(omp_priv = INIT)
\end{lstlisting}
\lstset{basicstyle=\small\ttfamily}
It involves casts of \lsttable{omp_in} and \lsttable{omp_out} to a \lstinline{Reducer} object that is associated with a global nested lock. Note that a nested lock has the distinguished property that it can be set multiple times from within the same thread without causing deadlocks. We implement the \lstinline{Reducer} such that the nested lock is set in constructor calls, which happens three times in the course of one reduction statement. The three set operations are complemented by three unset operations as part of the overload of \lsttable{operator =} for the \lsttable{Reducer}. This ensures that the whole reduction statement is exclusively performed by one thread at a time. This produces a serialization of the reduction tree in the forward pass that is preserved by the mutex AD handling for the reverse pass. In addition, we ensure via templating techniques that different custom reductions use different \lsttable{Reducer} objects and hence, different nested locks.

\section{Performance Study}
\label{section:performance_study}

To study the performance of OpDiLib, we apply it to a PDE solver for the coupled Burgers' equations
\begin{equation}
\begin{aligned}
u_t+uu_x+vu_y&=\frac{1}{R}(u_{xx}+u_{yy}),\\
v_t+uv_x+vv_y&=\frac{1}{R}(v_{xx}+v_{yy})
\end{aligned}
\label{eq:burgers}
\end{equation}
\cite{Bahadir2003,BiazarA2009,Zhu2010} that was previously used for the performance evaluation of various aspects of automatic differentiation in \cite{SagebaumAG2018,SagebaumAG2019,SagebaumBG2020}. We solve \eqref{eq:burgers} on $(x,\,y,\,t)\in D\times\RR_{\geq 0}$ where $D$ is a scaling of the unit square $[0,\,1]^2$, together with boundary conditions and initial conditions at $t=0$ which are taken from the exact solution
\begin{equation*}
\begin{aligned}
u(x,\,y,\,t)&=\frac{x+y-2xt}{1-2t^2},\\
v(x,\,y,\,t)&=\frac{x-y-2yt}{1-2t^2}
\end{aligned}
\end{equation*}
that is given in \cite{BiazarA2009}. We solve the equations with an upwind finite difference scheme, for which we discretize $D$ by an equidistant grid and advance in time steps of fixed size $\Delta t$. The simulations in this paper were carried out with $R=1$, $D=[0,50]^2$, on a grid with $2000\times 2000$ cells and for 20 time steps of width $\Delta t=0.0001$.
We use the reverse mode of AD to differentiate the $L^2$ norm of the solution after the final time step with respect to the initial data. CoDiPack is used as the underlying AD tool, configured for Jacobian taping \cite{Hogan2014} with a reuse index handler \cite{SagebaumBG2020}. For the differentiation of OpenMP, we couple it with OpDiLib as explained in Section \ref{section:ad_tool}. CoDiPack is also coupled with MeDiPack for the differentiation of MPI.
We use \verb|hwloc|\footnote{\url{https://www.open-mpi.org/projects/hwloc/}} for appropriate binding of MPI processes, and tie OpenMP threads to cores using OpenMP environment variables \citepalias{OpenMP52}.
Dynamic frequency scaling is disabled to improve the consistency of the performance results.
We display results obtained with the compilers \verb|clang++| 10.0.1\footnote{\url{https://releases.llvm.org/}} and \verb|g++| 10.1\footnote{\url{https://gcc.gnu.org/gcc-10/}}. The latter does not support OMPT yet and is hence an important use case for the macro backend presented in Section \ref{section:macro_backend}.

\begin{figure}
\includegraphics[width=\textwidth]{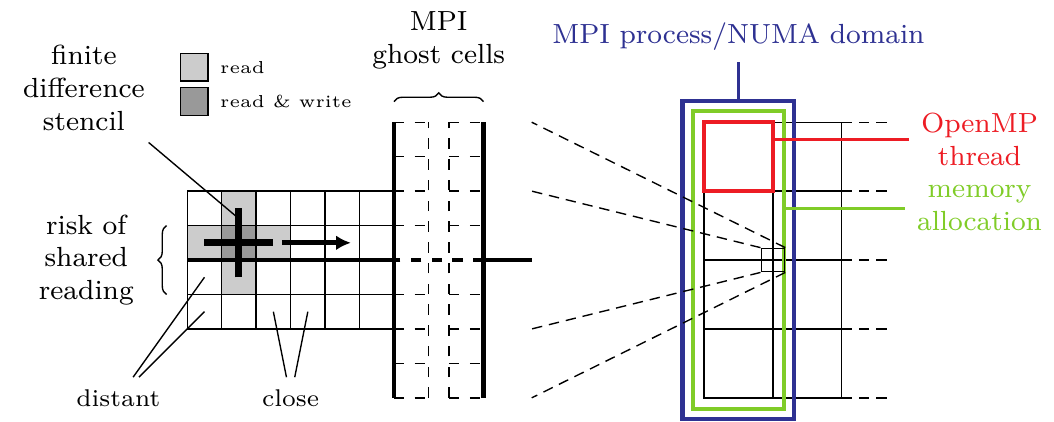}
\caption{Domain decomposition strategy with one MPI process per NUMA domain (right). We refer to this as \emph{NUMA splitting}. Memory is allocated per MPI process. The detailed view of the boundaries between MPI processes and OpenMP threads (left) also indicates non-exclusive read access at the OpenMP block boundaries as well as close and distant locations in the row-major memory layout.}
\label{figure:numa_layout}
\end{figure}

\begin{figure}
\begin{minipage}[t]{\textwidth}
	\centering
  \includegraphics[width=0.88\textwidth]{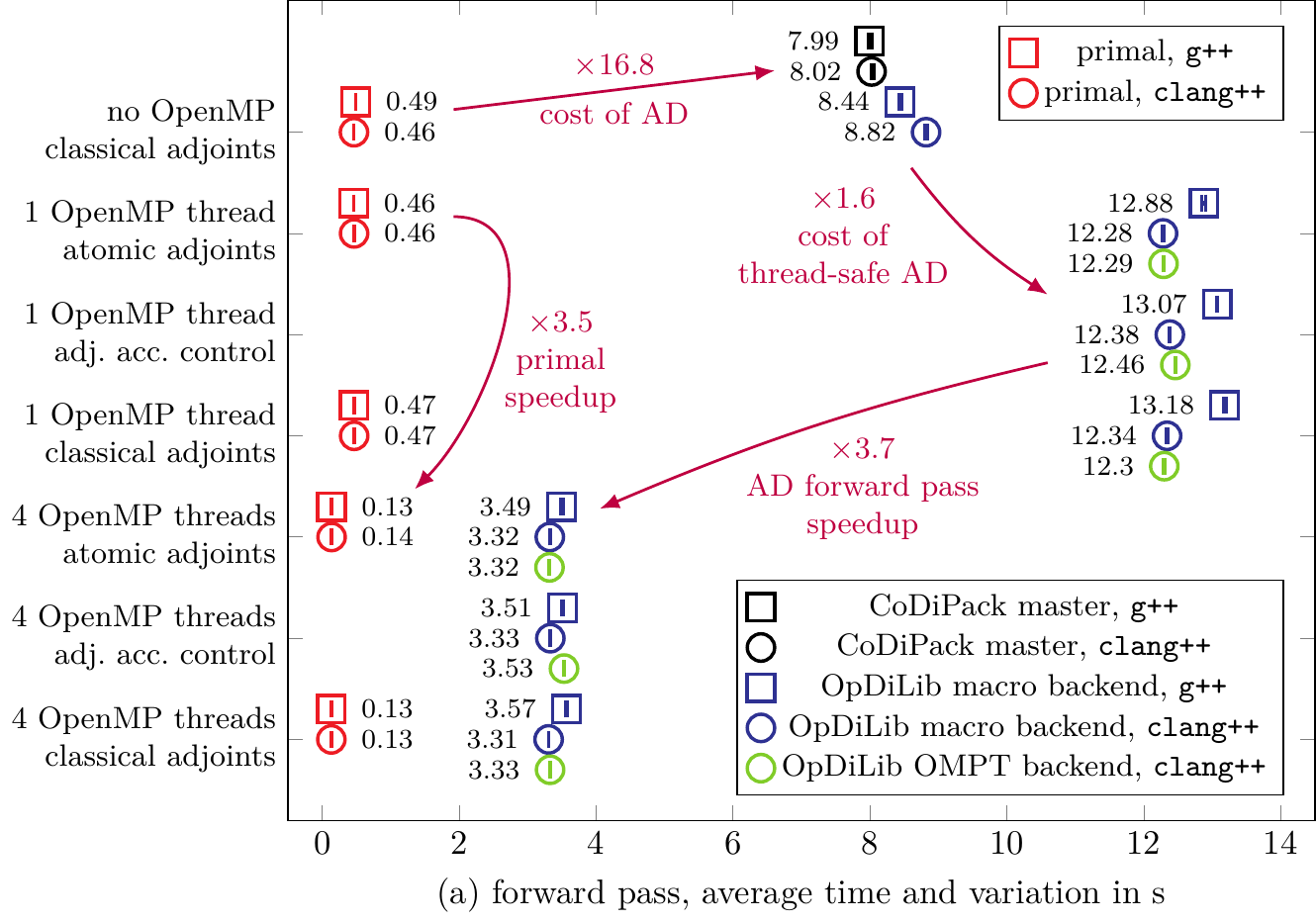}
\end{minipage}

\vspace{0.05cm}
\begin{minipage}[t]{\textwidth}
	\centering
  \includegraphics[width=0.88\textwidth]{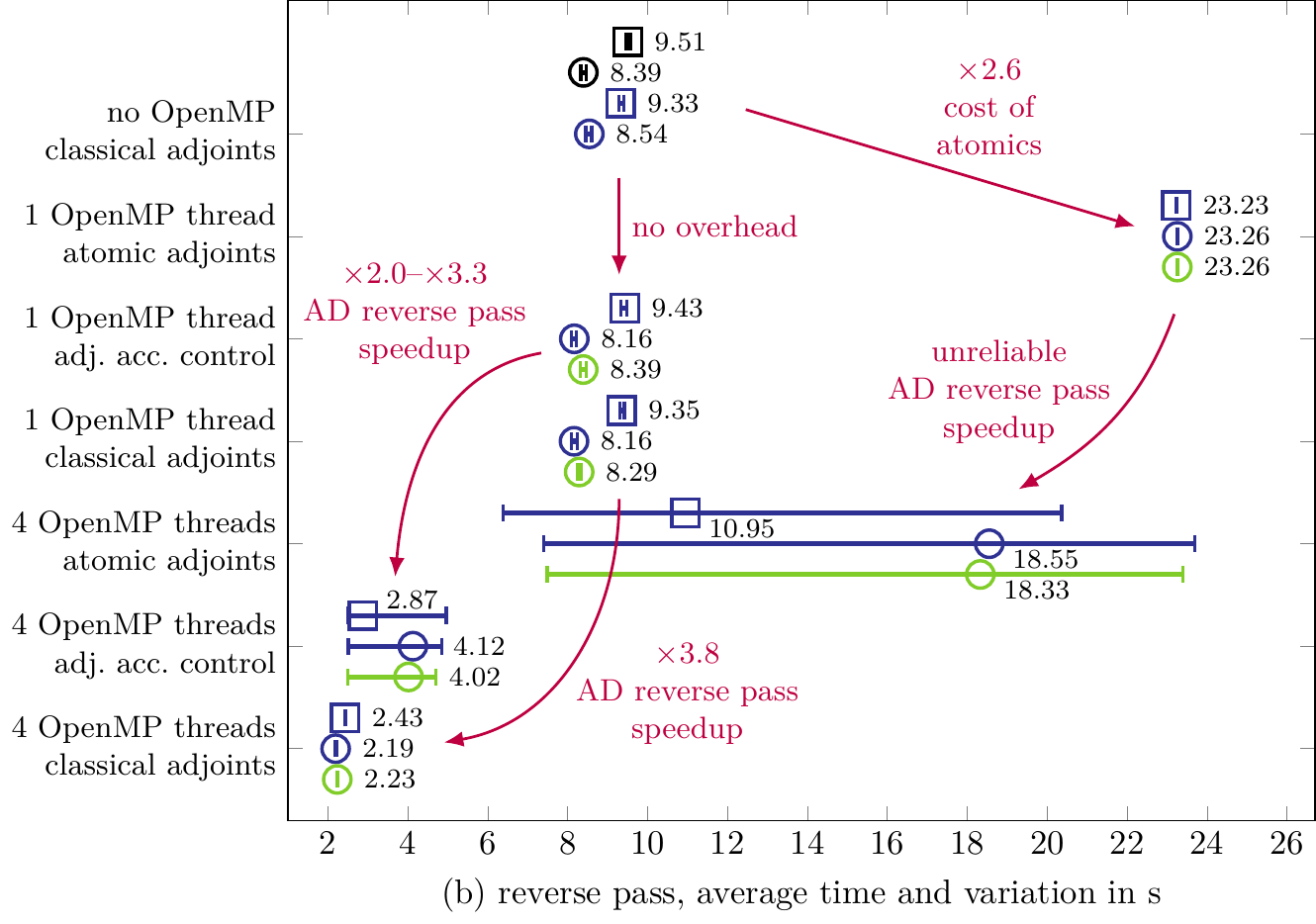}
\end{minipage}
\caption{Forward pass (a) and reverse pass (b) performance comparison of different parallel configurations within one NUMA domain of the AMD EPYC 7351 processor, consisting of 4 cores. Includes setups with \textcolor{black}{\textbf{classical AD}}, with the \textcolor{Blue}{\textbf{macro backend}} and with the \textcolor{LimeGreen}{\textbf{OMPT backend}}, compiled with \texttt{g++} \Large$\square$\small{} and \texttt{clang++} \raisebox{1pt}{$\bigcirc$}. In (a), we also display \textcolor{red}{\textbf{primal}} performance values. In the primal versions, atomic adjoints and adjoint access control configurations coincide. Numbers denote averages of 20 runs after five discarded warm-up runs. \textcolor{purple}{\textbf{Arrows}} indicate average cost and speedup relations.}
\label{figure:epyc_performance_comparison}
\end{figure}

We begin with a detailed performance study on a dual-socket node of the Elwetritsch cluster at TU Kaiserslautern with two AMD EPYC 7351 processors at 2.4\,GHz. Each processor consists of four NUMA domains with four cores each, giving a total of 32 cores per node. We use distributed memory parallelization between NUMA domains and employ shared memory parallelization within them. Solution fields are allocated per MPI process. Since adjoint variables are also allocated per MPI process, the allocation pattern and memory granularity with respect to NUMA domains are identical in the forward and reverse pass.
Figure \ref{figure:numa_layout} visualizes the corresponding hybrid parallel splitting strategy. The spatial domain is first decomposed into columns of equal width, one for each MPI process. Each MPI process stores its part of the solution field in row-major order, with padding so that the beginnings of rows are aligned with cache lines. Halo exchange via MPI communication is required at the column boundaries. Second, each MPI process works on its column with multiple OpenMP threads. To that end, the column is horizontally subdivided into blocks of equal height, resulting in one block for each OpenMP thread on which it performs row-wise sweeps with the five-point finite difference stencil.
As can be seen in Figure \ref{figure:numa_layout}, shared reading might occur at the boundary between adjacent blocks assigned to different OpenMP threads. The issue of atomic updates discussed in Section \ref{subsec:data_access_patterns} is reflected in the performance tests by three different adjoint configurations. While atomics are only used in the reverse pass, the respective code annotations and data access pattern adaptions also affect the forward pass.
\begin{description}
\item[Atomic adjoints] All updates on all adjoint variables are performed atomically. There are no restrictions on data access patterns in the forward pass. No code modifications are needed.
\item[Adjoint access control] The code is annotated with the adjoint access control tools presented in Section \ref{subsec:data_access_patterns}. We default to the \lstinline{Classical} adjoint access mode. Whenever we process the two rows adjacent to the neighbouring block, we switch to the \lstinline{Atomic} adjoint access mode.
\item[Classical adjoints] No atomics are used in the reverse pass. To guarantee exclusive read access in the forward pass, we double the number of blocks per MPI column and process them with the OpenMP threads in two sweeps in an interleaved manner. A reverse barrier is required between the two sweeps.
\end{description}
We start with a comparison of absolute performance values of the different adjoint configurations within one NUMA domain of the AMD EPYC 7351 processor, in which we execute the simulation with either one or four OpenMP threads. We also compare the performance to primal, undifferentiated and classical, serially differentiated versions of the simulation. The results are displayed in Figure \ref{figure:epyc_performance_comparison}.
The different configurations are denoted on the vertical axis whereas the timings obtained with different compilers and AD setups are denoted on the horizontal axis. We begin with some general observations. If we compare for each configuration the impact of the different compilers (squares versus circles) and the impact of the backend choice (blue versus green), we see that neither of both has a significant influence on the performance. Furthermore, the adjoint configurations do not have an impact on the performance of the forward pass. The OpDiLib results obtained without OpenMP still make use of the thread-safe version of CoDiPack and hence, allow us to evaluate the impact of the CoDiPack modifications on the performance of the standard CoDiPack type (black versus blue). We see that it is mostly preserved, especially in the reverse pass. Using one OpenMP thread, we see a certain performance overhead in the forward pass that is not visible in the primal performance values. The major part of this is due to using a thread-safe CoDiPack type, with minor parts due to using OpenMP and OpDiLib. In the reverse pass, there are significant differences between the three adjoint configurations. While the adjoint access control approach and the classical adjoints strategy fully reproduce the performance of the classical, serially differentiated runs, using atomic adjoints incurs a significant performance overhead. With four OpenMP threads, the speedup of the AD forward pass is similar to the speedup of the primal, undifferentiated simulation, even slightly larger, regardless of the adjoint configuration. In terms of AD, this is also an improvement of more than a factor of two compared to the classical, serially differentiated versions. A parallel reverse pass with atomic adjoints is not only notably slower on average than the classical, serial reverse run but also exhibits huge variation in the timings, resulting in the most unreliable performance among all adjoint configurations. With the adjoint access control approach, the parallel reverse pass overhead is reduced significantly. This shows that if there are only a few variables without exclusive read access, the adjoint access control tools provide a way to achieve acceptable parallel reverse performance without revising the data access patterns. The most reliable reverse pass speedup that is consistently at par with the forward pass speedup is obtained with the classical adjoints strategy, that does not use atomic updates at all. Among the atomic adjoints and adjoint access control configurations, the displayed test results seem to suggest differences between compilers and OpDiLib backends both with respect to the extent of noise and the average performance, but this cannot be reproduced in a reliable manner and admits no conclusion. We remark that there is almost no difference between the memory high water marks of the tested AD setups. Compared to the classical, serially differentiated simulation, any additional memory consumption due to OpDiLib's differentiation logic and backend is negligible.

\begin{figure}
\centering
\includegraphics[width=1\textwidth]{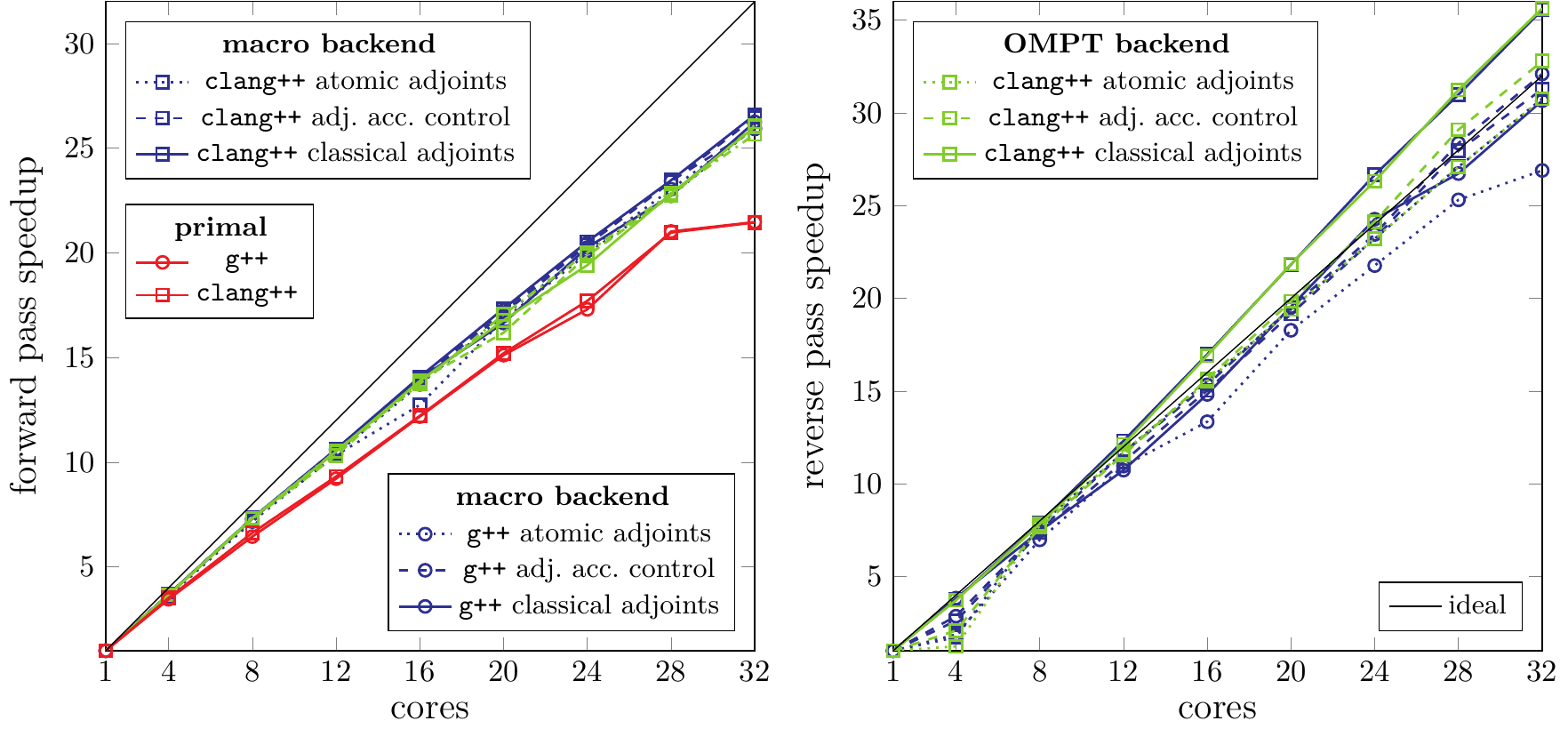}
\caption{Scaling results of OpDiLib on a dual-socket AMD EPYC 7351 node for the forward pass (left) and the reverse pass (right). Includes primal scaling results in the forward pass and the ideal scaling curves. Beyond the baseline, the number of OpenMP threads per MPI process is fixed to 4 and the number of MPI processes is increased up to 8 (one per NUMA domain).}
\label{figure:epyc_scaling}
\end{figure}

Next, we investigate the strong scaling of the adjoint configurations. In the forward pass, we also include the scaling of the primal, undifferentiated code. The scaling strategy is implied by the NUMA splitting displayed in Figure \ref{figure:numa_layout} together with the hardware characteristics of the AMD EPYC 7351 processor. For each configuration, the timings obtained with one MPI process and one OpenMP thread serve as the baseline. We first increase the number of OpenMP threads per MPI process to four, and then increase the number of MPI processes until the full node is used. The results are displayed in Figure \ref{figure:epyc_scaling}.
We see that all AD setups scale better than the primal, undifferentiated simulation. This is often observed in AD and is attributed to the fact that AD improves the computation to memory ratio in the forward pass \cite{NemiliOGKT2016}. Neither the OpDiLib backend choice nor the compiler nor the adjoint configuration have a significant influence on the scaling in the forward pass. In the reverse pass, the OpDiLib backend choice does not have a significant impact either. \verb|clang++| achieves better scaling in the reverse pass, and less atomics correlate with better scaling. Overall, the reverse scaling results are very good, some configurations exhibit even superlinear behaviour, which is fostered by multiple factors. In the splitting strategy displayed in Figure \ref{figure:numa_layout}, shared reading --- and hence serialization in the reverse pass --- can only occur at the cells at block boundaries between adjacent OpenMP threads. The number of these cells is furthermore invariant with respect to the number of MPI processes so that, in the worst case, there is a constant small fraction of the computational tasks that does not scale. Furthermore, the splitting strategy itself supports scaling. Increasing the number of MPI processes corresponds to decreasing the width of the column for each process. This, in turn, implies that the distant memory locations that are accessed during one application of the finite difference stencil move closer to each other, see Figure \ref{figure:numa_layout}. This applies analogously to the corresponding pattern of adjoint variable access in the reverse pass. Hence, memory locality improves with an increasing number of MPI processes.

\begin{figure}
\includegraphics[width=\textwidth]{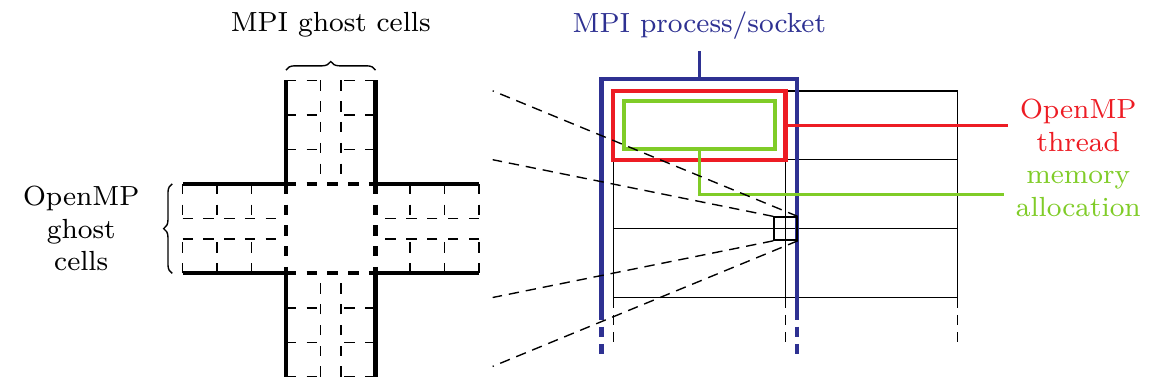}
\caption{Domain decomposition strategy with one MPI process per socket (right). We refer to this as \emph{socket splitting}. As indicate in the detailed view of the boundaries between MPI processes and OpenMP threads (left), OpenMP communication uses halo exchange, with which exclusive read access is guaranteed.}
\label{figure:socket_layout}
\end{figure}

\begin{figure}
\centering
\includegraphics[width=1\textwidth]{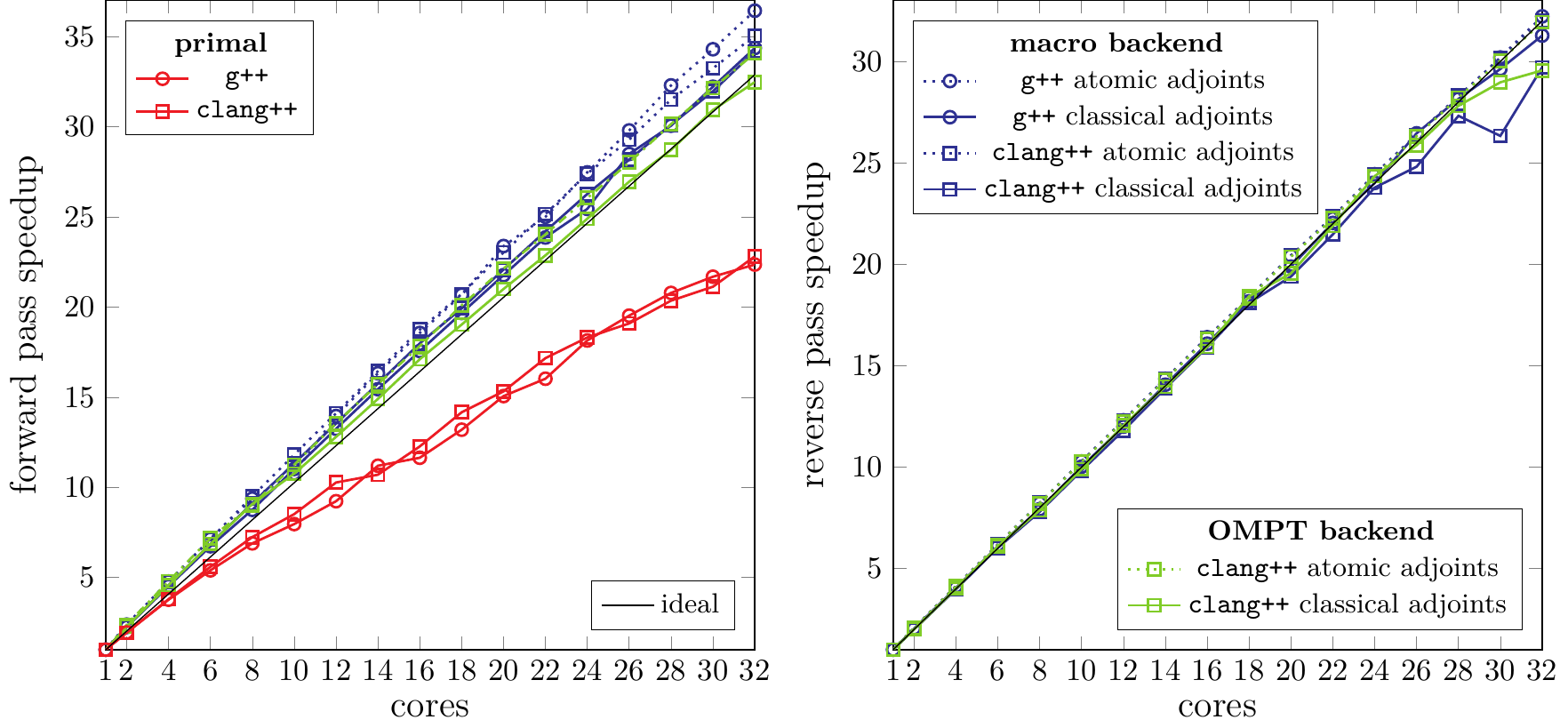}
\caption{Scaling results of OpDiLib on a dual-socket AMD EPYC 7351 node for the forward pass (left) and the reverse pass (right). Includes primal scaling results in the forward pass and the ideal scaling curves. Beyond the baseline, the number MPI processes is fixed to two (one per socket) and the number of OpenMP threads per MPI process is increased up to 16.}
\label{figure:epyc_scaling_socket}
\end{figure}

We complement the performance results obtained so far by scaling tests with a more OpenMP centered approach. To that end, we use one MPI process per socket with up to 16 OpenMP threads on the AMD EPYC 7351 processor. To better distribute the memory load across the NUMA domains, we move from memory allocations per MPI process to memory allocations per thread. This results in a new splitting strategy that is visualized in Figure \ref{figure:socket_layout}.
Neighbouring threads no longer access each other's memory directly. Instead, we use halo exchange analogous to MPI communication, but implemented with OpenMP. This involves communication buffers and suitable synchronization; as a side effect, exclusive read access is guaranteed in the forward pass for all variables, which in turn impacts the adjoint configurations. Even though no atomic updates are needed, we can use them nonetheless to investigate their impact on the performance. Adjoint access control, however, is not meaningful in  in the scope of this splitting strategy.
The results are displayed in Figure \ref{figure:epyc_scaling_socket}.
They demonstrate that very good scaling can be achieved with OpDiLib also with larger numbers of OpenMP threads per MPI process.

\begin{figure}
\centering
\includegraphics[width=1\textwidth]{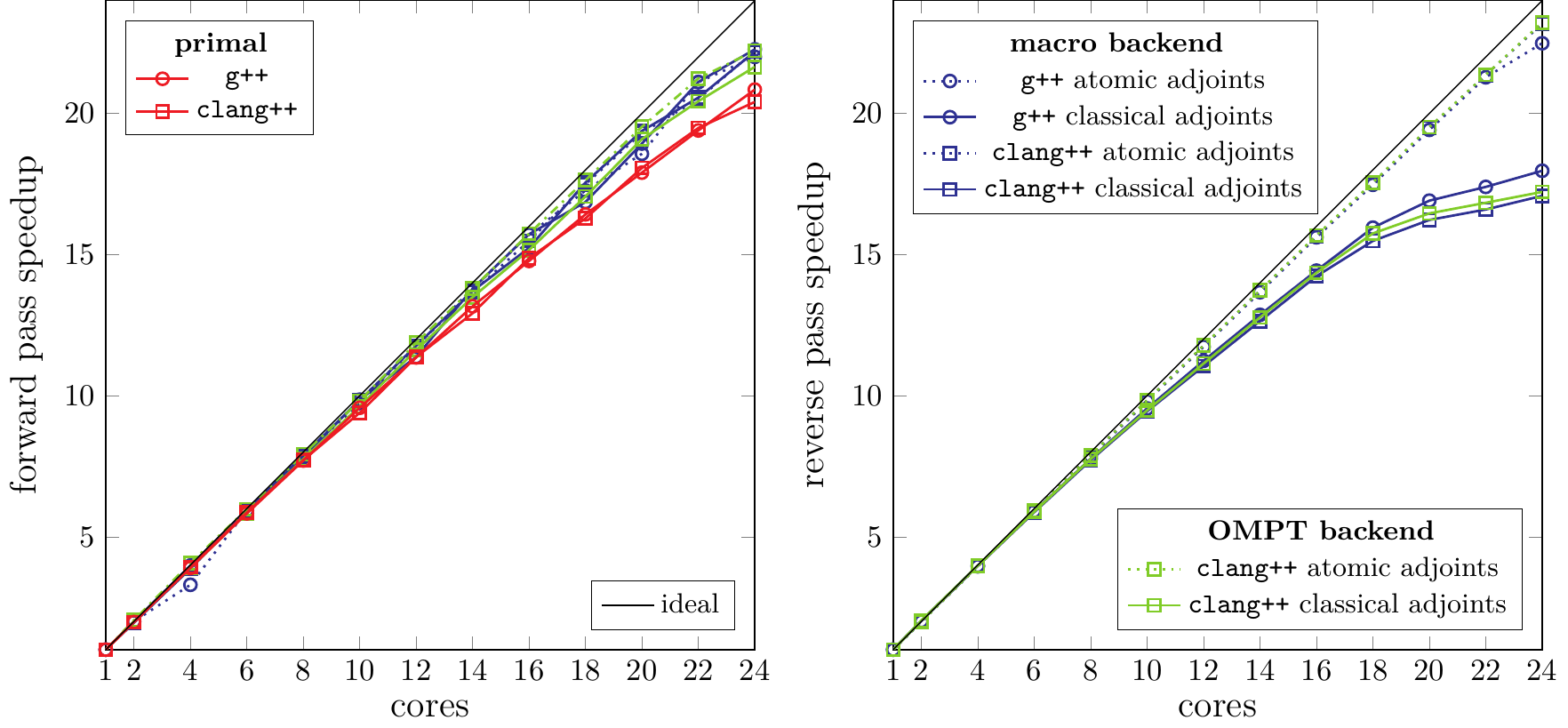}
\caption{Scaling results of OpDiLib on a dual-socket Intel Xeon Gold 6126 node for the forward pass (left) and the reverse pass (right). Includes primal scaling results in the forward pass and the ideal scaling curves. Beyond the baseline, the number of MPI processes is fixed to two (one per socket) and the number of OpenMP threads per MPI process is increased up to 12.}
\label{figure:skylake_scaling_socket}
\end{figure}

\begin{figure}
\centering
\includegraphics[width=1\textwidth]{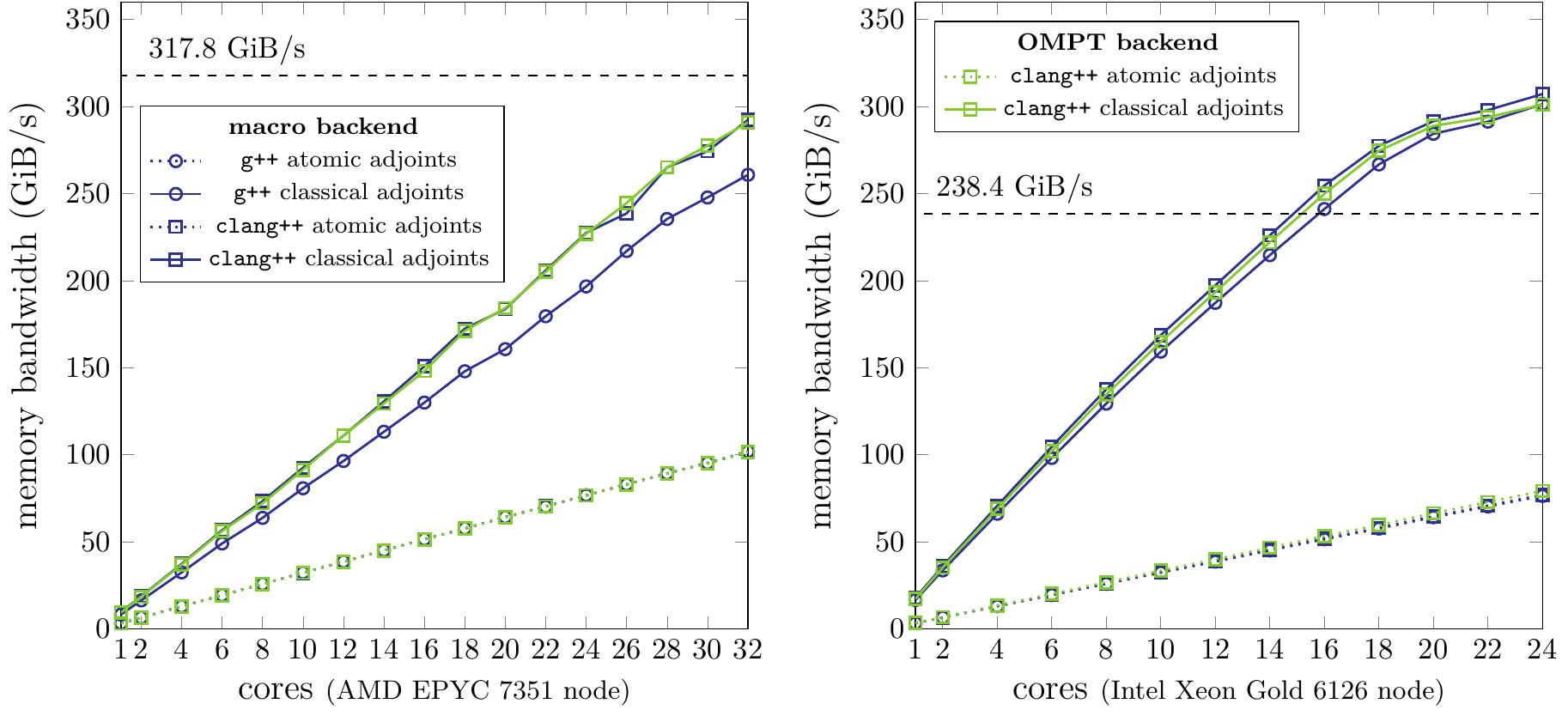}
\caption{Memory bandwidth achieved during the reverse pass with the per-socket splitting approach on the AMD EPYC 7351 node (left) and the Intel Xeon Gold 6126 node (right). Bandwidth is computed from the measured runtime, the  tape sizes, and the known load and store operations for evaluating statements. Both plots indicate the maximum memory bandwidth of the respective node. Achieved bandwidth can be higher due to caching.}
\label{figure:bandwidth}
\end{figure}

The reverse pass in tape-based AD is memory-bound, also if OpDiLib is used. We can make this visible on a different type of testing hardware, a dual-socket node of the Elwetritsch cluster with two Intel Xeon Gold 6126 processors at 2.6\,GHz. Each processor consists of 12 cores that form a single NUMA domain. As we repeat the previous tests with the two splitting approaches adapted to the different NUMA layout and core number, the trends observed in Figure \ref{figure:epyc_performance_comparison} are reproduced, but configurations with atomics perform less noisy and the individual cost and speedup factors are different.
Particularly, the overall speedup factors are smaller relative to the number of used cores, even though this can be improved by moving to the socket splitting strategy with memory allocations per thread from Figure \ref{figure:socket_layout}. An overview over all results with the OMPT backend and \texttt{clang++} with both splitting approaches and both types of processors is given in Table \ref{table:ompt_all_results}.
\begin{table}
\centering
\caption{Forward pass results (top) and reverse pass results (bottom) with the OMPT backend and \texttt{clang++} on the AMD EPYC 7351 node as well as the Intel Xeon Gold 6126 node, with the NUMA splitting approach from Figure \protect{\ref{figure:numa_layout}} and the socket splitting approach from Figure \protect{\ref{figure:socket_layout}}. For better comparability, we provide the parallel efficiency, that is, speedup normalized by the number of cores. Includes primal results in the forward pass.}
\label{table:ompt_all_results}
\small
\begin{tabular}{rcccccc}
\toprule
forward pass & \multicolumn{3}{c}{one NUMA domain, NUMA splitting} & \multicolumn{3}{c}{one socket, socket splitting} \\
\midrule
AMD EPYC 7351 & \makecell[c]{1 thread,\\time in s} & \makecell[c]{4 threads,\\time in s} & \makecell[c]{parallel\\efficiency} & \makecell[c]{1 thread,\\time in s} & \makecell[c]{16 threads,\\time in s} & \makecell[c]{parallel\\efficiency} \\
\midrule
primal & 0.46 & 0.14 & 0.85 & 0.46 & 0.04 & 0.73\\
atomic adjoints & 12.29 & 3.32 & 0.93 & 14.96 & 0.83 & 1.12 \\
adjoint access control & 12.46 & 3.53 & 0.88 & -- & -- & --  \\
classical adjoints & 12.30 & 3.33 & 0.92 & 14.17 & 0.85 & 1.04  \\
\midrule
Intel Xeon Gold 6126 & \makecell[c]{1 thread,\\time in s} & \makecell[c]{12 threads,\\time in s} & \makecell[c]{parallel\\efficiency} & \makecell[c]{1 thread,\\time in s} & \makecell[c]{12 threads,\\time in s} & \makecell[c]{parallel\\efficiency} \\
\midrule
primal & 0.68 & 0.08 & 0.74 & 0.68 & 0.06 & 0.94\\
atomic adjoints & 13.03 & 1.34 & 0.81 & 13.16 & 1.17 & 0.94 \\
adjoint access control & 13.10 & 1.38 & 0.79 & -- & -- & -- \\
classical adjoints & 13.08 & 1.34 & 0.81 & 13.15 & 1.17 & 0.94 \\
\bottomrule
\end{tabular}
\begin{tabular}{rcccccc}
\toprule
reverse pass & \multicolumn{3}{c}{one NUMA domain, NUMA splitting} & \multicolumn{3}{c}{one socket, socket splitting} \\
\midrule
AMD EPYC 7351 & \makecell[c]{1 thread,\\time in s} & \makecell[c]{4 threads,\\time in s} & \makecell[c]{parallel\\efficiency} & \makecell[c]{1 thread,\\time in s} & \makecell[c]{16 threads,\\time in s} & \makecell[c]{parallel\\efficiency} \\
\midrule
atomic adjoints & 23.26 & 18.33 & 0.32 & 24.26 & 1.50 & 1.01 \\
adjoint access control & 8.39 & 4.02 & 0.52 & -- & -- & --  \\
classical adjoints & 8.29 & 2.23 & 0.93 & 8.65 & 0.55 & 0.98  \\
\midrule
Intel Xeon Gold 6126 & \makecell[c]{1 thread,\\time in s} & \makecell[c]{12 threads,\\time in s} & \makecell[c]{parallel\\efficiency} & \makecell[c]{1 thread,\\time in s} & \makecell[c]{12 threads,\\time in s} & \makecell[c]{parallel\\efficiency} \\
\midrule
atomic adjoints & 22.62 & 4.45 & 0.42 & 22.43 & 1.94 & 0.96 \\
adjoint access control & 4.24 & 0.78 & 0.45 & -- & -- & -- \\
classical adjoints & 4.28 & 0.54 & 0.66 & 4.39 & 0.53 & 0.70 \\
\bottomrule
\end{tabular}
\end{table}
Figure \ref{figure:skylake_scaling_socket} shows the scaling observed with socket splitting on the Intel Xeon Gold 6126 node.
Clearly, configurations without atomic updates plateau whereas configurations with atomics keep scaling up to the full node. The reason is that the first operate closer to the memory bandwidth, and become limited by it. The relation to the memory bandwidth utilization in the reverse pass is investigated for both the AMD EPYC 7351 node and the Intel Xeon Gold 6126 node in Figure \ref{figure:bandwidth}.
As expected from the previous scaling results, the bandwidth limit is not exceeded on the AMD node. On the Intel node, speedup beyond the hardware limit is only possible due to caching. Consequently, both the bandwidth and the scaling curves flatten.

Overall, the test results show that very good parallel AD performance can be achieved with OpDiLib. High parallel efficiencies show that OpDiLib succeeds at scaling serial tape-based AD to multiple OpenMP threads. The overall performance is strongly influenced by hardware characteristics, which can be addressed by adapting data decomposition and parallelization accordingly. The reverse pass timings are primarily influenced by atomic updates. Depending on the application, confining these to the variables that actually need them might already yield the desired speedup.

\section{Summary and Conclusion}
\label{section:conclusion}

With OpDiLib, we have designed a tool for the automatic differentiation of OpenMP parallel codes that is fast, flexible, reusable and easy to apply. Our interpretation of AD augmentations as events alongside the OpenMP workflow allowed us to cover an unprecedented extent of OpenMP mechanisms in an operator overloading AD tool. We provide this functionality in the new open source software OpDiLib
as a universal add-on for operator overloading AD tools. We publish it together with bindings for CoDiPack and hope that it is also a useful addition to other operator overloading AD tools. In order to couple OpDiLib to another AD tool, that tool must be made thread-safe in a first step. The coupling performed in the second step provides capabilities for handling OpenMP constructs including an automatic deduction of a parallel reverse pass. With the OMPT backend, we achieve full automatization, that is, it does not require additional code changes. We do not restrict the data access patterns either. Hence, if a serial code that was already differentiated with the classical AD tool is parallelized by means of OpenMP, an initial parallel differentiated code is obtained immediately.
While we apply the conservative strategy of atomic updates on the adjoint variables, we do not prevent further optimization and equip the user with the option to disable atomic evaluation for any parts of the code. We complemented the OMPT features by an alternative macro backend, which serves as a fallback solution and also covers compilers without OMPT support. OpDiLib's flexible design allowed us to support the different use cases and modes of operation presented in this paper and will likely benefit future modifications and applications.
We have demonstrated in the performance tests that all configurations of OpDiLib preserve the serial AD memory footprint and that very good parallel performance can be achieved in practice.
We will enhance and improve OpDiLib in the future and work on additional features and configurations is in progress. OpDiLib's OpenMP support as presented in this paper covers all directives, clauses and runtime functions of the OpenMP 2.5 specification \citepalias{OpenMP25} with the exception of atomics and the flush directive, as we explained in Section \ref{subsec:transformation_of_openmp}. This resembles the core feature set of OpenMP for classical CPU architectures. We additionally demonstrated OpDiLib's applicability for the pure operator overloading automatic differentiation of OpenMP-MPI hybrid parallel codes. Hence, already now, OpDiLib should be of interest to many simulation codes executed on CPU computing clusters.

\bibliographystyle{ACM-Reference-Format}
\bibliography{literature}{}


\begin{thebibliography}{39}


\ifx \showCODEN    \undefined \def \showCODEN     #1{\unskip}     \fi
\ifx \showDOI      \undefined \def \showDOI       #1{#1}\fi
\ifx \showISBNx    \undefined \def \showISBNx     #1{\unskip}     \fi
\ifx \showISBNxiii \undefined \def \showISBNxiii  #1{\unskip}     \fi
\ifx \showISSN     \undefined \def \showISSN      #1{\unskip}     \fi
\ifx \showLCCN     \undefined \def \showLCCN      #1{\unskip}     \fi
\ifx \shownote     \undefined \def \shownote      #1{#1}          \fi
\ifx \showarticletitle \undefined \def \showarticletitle #1{#1}   \fi
\ifx \showURL      \undefined \def \showURL       {\relax}        \fi
\providecommand\bibfield[2]{#2}
\providecommand\bibinfo[2]{#2}
\providecommand\natexlab[1]{#1}
\providecommand\showeprint[2][]{arXiv:#2}

\bibitem[Auroux and Groza(2017)]%
        {AurouxG2017}
\bibfield{author}{\bibinfo{person}{D. Auroux} {and} \bibinfo{person}{V.
  Groza}.} \bibinfo{year}{2017}\natexlab{}.
\newblock \showarticletitle{Optimal parameters identification and sensitivity
  study for abrasive waterjet milling model}.
\newblock \bibinfo{journal}{\emph{{Inverse Probl. Sci. Eng.}}}
  \bibinfo{volume}{25}, \bibinfo{number}{11} (\bibinfo{date}{Dec.}
  \bibinfo{year}{2017}), \bibinfo{pages}{1560--1576}.
\newblock
\urldef\tempurl%
\url{https://doi.org/10.1080/17415977.2016.1273916}
\showDOI{\tempurl}


\bibitem[Bahadır(2003)]%
        {Bahadir2003}
\bibfield{author}{\bibinfo{person}{A.~Refik Bahadır}.}
  \bibinfo{year}{2003}\natexlab{}.
\newblock \showarticletitle{A fully implicit finite-difference scheme for
  two-dimensional {B}urgers' equations}.
\newblock \bibinfo{journal}{\emph{Appl. Math. Comput.}} \bibinfo{volume}{137},
  \bibinfo{number}{1} (\bibinfo{date}{May} \bibinfo{year}{2003}),
  \bibinfo{pages}{131--137}.
\newblock
\urldef\tempurl%
\url{https://doi.org/10.1016/S0096-3003(02)00091-7}
\showDOI{\tempurl}


\bibitem[Baydin et~al\mbox{.}(2017)]%
        {BaydinPRS2017}
\bibfield{author}{\bibinfo{person}{At\i{}l\i{}m~G\"{u}nes Baydin},
  \bibinfo{person}{Barak~A. Pearlmutter}, \bibinfo{person}{Alexey~Andreyevich
  Radul}, {and} \bibinfo{person}{Jeffrey~Mark Siskind}.}
  \bibinfo{year}{2017}\natexlab{}.
\newblock \showarticletitle{Automatic differentiation in machine learning: a
  survey}.
\newblock \bibinfo{journal}{\emph{J. Mach. Learn. Res.}} \bibinfo{volume}{18},
  \bibinfo{number}{153} (\bibinfo{date}{Jan.} \bibinfo{year}{2017}),
  \bibinfo{pages}{1--43}.
\newblock
\urldef\tempurl%
\url{https://jmlr.org/papers/v18/17-468.html}
\showURL{%
\tempurl}


\bibitem[Biazar and Aminikhah(2009)]%
        {BiazarA2009}
\bibfield{author}{\bibinfo{person}{J. Biazar} {and} \bibinfo{person}{H.
  Aminikhah}.} \bibinfo{year}{2009}\natexlab{}.
\newblock \showarticletitle{Exact and numerical solutions for non-linear
  {B}urger's equation by {VIM}}.
\newblock \bibinfo{journal}{\emph{Math. Comput. Modell.}} \bibinfo{volume}{49},
  \bibinfo{number}{7} (\bibinfo{date}{Apr.} \bibinfo{year}{2009}),
  \bibinfo{pages}{1394--1400}.
\newblock
\urldef\tempurl%
\url{https://doi.org/10.1016/j.mcm.2008.12.006}
\showDOI{\tempurl}


\bibitem[Bischof et~al\mbox{.}(2008)]%
        {BischofGKW2008}
\bibfield{author}{\bibinfo{person}{Christian Bischof}, \bibinfo{person}{Niels
  Guertler}, \bibinfo{person}{Andreas Kowarz}, {and} \bibinfo{person}{Andrea
  Walther}.} \bibinfo{year}{2008}\natexlab{}.
\newblock \showarticletitle{Parallel reverse mode automatic differentiation for
  {OpenMP} programs with {ADOL-C}}. In \bibinfo{booktitle}{\emph{Advances in
  Automatic Differentiation}}, \bibfield{editor}{\bibinfo{person}{Christian~H.
  Bischof}, \bibinfo{person}{H.~Martin B{\"u}cker}, \bibinfo{person}{Paul
  Hovland}, \bibinfo{person}{Uwe Naumann}, {and} \bibinfo{person}{Jean Utke}}
  (Eds.). \bibinfo{publisher}{Springer}, \bibinfo{address}{Berlin, Heidelberg,
  Germany}, \bibinfo{pages}{163--173}.
\newblock
\urldef\tempurl%
\url{https://doi.org/10.1007/978-3-540-68942-3_15}
\showDOI{\tempurl}


\bibitem[B\"{u}cker et~al\mbox{.}(2001)]%
        {BueckerLMB2001}
\bibfield{author}{\bibinfo{person}{H.~Martin B\"{u}cker},
  \bibinfo{person}{Bruno Lang}, \bibinfo{person}{Dieter an Mey}, {and}
  \bibinfo{person}{Christian~H. Bischof}.} \bibinfo{year}{2001}\natexlab{}.
\newblock \showarticletitle{Bringing together automatic differentiation and
  {OpenMP}}. In \bibinfo{booktitle}{\emph{Proceedings of the 15th International
  Conference on Supercomputing}} (Sorrento, Italy). \bibinfo{publisher}{ACM},
  \bibinfo{address}{New York, NY, USA}, \bibinfo{pages}{246--251}.
\newblock
\urldef\tempurl%
\url{https://doi.org/10.1145/377792.377842}
\showDOI{\tempurl}


\bibitem[B\"{u}cker et~al\mbox{.}(2002)]%
        {BueckerLRBM2002}
\bibfield{author}{\bibinfo{person}{H.~Martin B\"{u}cker},
  \bibinfo{person}{Bruno Lang}, \bibinfo{person}{Arno Rasch},
  \bibinfo{person}{Christian~H. Bischof}, {and} \bibinfo{person}{Dieter an
  Mey}.} \bibinfo{year}{2002}\natexlab{}.
\newblock \showarticletitle{Explicit loop scheduling in {OpenMP} for parallel
  automatic differentiation}. In \bibinfo{booktitle}{\emph{Proceedings 16th
  Annual International Symposium on High Performance Computing Systems and
  Applications}} (Moncton, NB, Canada). \bibinfo{publisher}{IEEE},
  \bibinfo{address}{New York, NY, USA}, \bibinfo{pages}{121--126}.
\newblock
\urldef\tempurl%
\url{https://doi.org/10.1109/HPCSA.2002.1019144}
\showDOI{\tempurl}


\bibitem[B\"{u}cker et~al\mbox{.}(2004)]%
        {BueckerRW2004}
\bibfield{author}{\bibinfo{person}{H.~Martin B\"{u}cker}, \bibinfo{person}{Arno
  Rasch}, {and} \bibinfo{person}{Andreas Wolf}.}
  \bibinfo{year}{2004}\natexlab{}.
\newblock \showarticletitle{A class of {OpenMP} applications involving nested
  parallelism}. In \bibinfo{booktitle}{\emph{Proceedings of the 2004 ACM
  Symposium on Applied Computing}} (Nicosia, Cyprus). \bibinfo{publisher}{ACM},
  \bibinfo{address}{New York, NY, USA}, \bibinfo{pages}{220--224}.
\newblock
\urldef\tempurl%
\url{https://doi.org/10.1145/967900.967948}
\showDOI{\tempurl}


\bibitem[Eichenberger et~al\mbox{.}(2013)]%
        {EichenbergerEtAl2013}
\bibfield{author}{\bibinfo{person}{Alexandre~E. Eichenberger},
  \bibinfo{person}{John Mellor-Crummey}, \bibinfo{person}{Martin Schulz},
  \bibinfo{person}{Michael Wong}, \bibinfo{person}{Nawal Copty},
  \bibinfo{person}{Robert Dietrich}, \bibinfo{person}{Xu Liu},
  \bibinfo{person}{Eugene Loh}, \bibinfo{person}{Daniel Lorenz},
  {et~al\mbox{.}}} \bibinfo{year}{2013}\natexlab{}.
\newblock \showarticletitle{{OMPT}: an {OpenMP} tools application programming
  interface for performance analysis}. In \bibinfo{booktitle}{\emph{{OpenMP in
  the Era of Low Power Devices and Accelerators}}},
  \bibfield{editor}{\bibinfo{person}{Alistair~P. Rendell},
  \bibinfo{person}{Barbara~M. Chapman}, {and} \bibinfo{person}{Matthias~S.
  M{\"u}ller}} (Eds.). \bibinfo{publisher}{Springer}, \bibinfo{address}{Berlin,
  Heidelberg, Germany}, \bibinfo{pages}{171--185}.
\newblock
\urldef\tempurl%
\url{https://doi.org/10.1007/978-3-642-40698-0_13}
\showDOI{\tempurl}


\bibitem[F{\"o}rster(2014)]%
        {Foerster2014}
\bibfield{author}{\bibinfo{person}{Michael F{\"o}rster}.}
  \bibinfo{year}{2014}\natexlab{}.
\newblock \bibinfo{booktitle}{\emph{{Algorithmic Differentiation of
  Pragma-Defined Parallel Regions}}}.
\newblock \bibinfo{publisher}{Springer Vieweg}, \bibinfo{address}{Wiesbaden,
  Germany}.
\newblock
\urldef\tempurl%
\url{https://doi.org/10.1007/978-3-658-07597-2}
\showDOI{\tempurl}
\newblock
\shownote{Ph.\,D. Dissertation. RWTH Aachen University, Aachen, Germany}.


\bibitem[Griewank and Walther(2008)]%
        {GriewankW2008}
\bibfield{author}{\bibinfo{person}{Andreas Griewank} {and}
  \bibinfo{person}{Andrea Walther}.} \bibinfo{year}{2008}\natexlab{}.
\newblock \bibinfo{booktitle}{\emph{{Evaluating Derivatives: Principles and
  Techniques of Algorithmic Differentiation}}}. Vol.~\bibinfo{volume}{105}.
\newblock \bibinfo{publisher}{SIAM}, \bibinfo{address}{Philadelphia, PA, USA}.
\newblock
\urldef\tempurl%
\url{https://doi.org/10.1137/1.9780898717761}
\showDOI{\tempurl}


\bibitem[G{\"u}nther et~al\mbox{.}(2020)]%
        {GuentherRSCG2020}
\bibfield{author}{\bibinfo{person}{Stefanie G{\"u}nther}, \bibinfo{person}{Lars
  Ruthotto}, \bibinfo{person}{Jacob~B. Schroder}, \bibinfo{person}{Eric~C.
  Cyr}, {and} \bibinfo{person}{Nicolas~R. Gauger}.}
  \bibinfo{year}{2020}\natexlab{}.
\newblock \showarticletitle{Layer-parallel training of deep residual neural
  networks}.
\newblock \bibinfo{journal}{\emph{SIAM J. Math. Data Sci.}}
  \bibinfo{volume}{2}, \bibinfo{number}{1} (\bibinfo{date}{Feb.}
  \bibinfo{year}{2020}), \bibinfo{pages}{1--23}.
\newblock
\urldef\tempurl%
\url{https://doi.org/10.1137/19M1247620}
\showDOI{\tempurl}


\bibitem[Hasco{\"e}t and Pascual(2013)]%
        {HascoetP2013}
\bibfield{author}{\bibinfo{person}{Laurent Hasco{\"e}t} {and}
  \bibinfo{person}{Val\'{e}rie Pascual}.} \bibinfo{year}{2013}\natexlab{}.
\newblock \showarticletitle{The {Tapenade} automatic differentiation tool:
  principles, model, and specification}.
\newblock \bibinfo{journal}{\emph{ACM Trans. Math. Softw.}}
  \bibinfo{volume}{39}, \bibinfo{number}{3}, Article \bibinfo{articleno}{20}
  (\bibinfo{date}{Apr.} \bibinfo{year}{2013}), \bibinfo{numpages}{43}~pages.
\newblock
\urldef\tempurl%
\url{https://doi.org/10.1145/2450153.2450158}
\showDOI{\tempurl}


\bibitem[Heimbach et~al\mbox{.}(2002)]%
        {HeimbachHG2002}
\bibfield{author}{\bibinfo{person}{Patrick Heimbach}, \bibinfo{person}{Chris
  Hill}, {and} \bibinfo{person}{Ralf Giering}.}
  \bibinfo{year}{2002}\natexlab{}.
\newblock \showarticletitle{Automatic generation of efficient adjoint code for
  a parallel {N}avier-{S}tokes solver}. In
  \bibinfo{booktitle}{\emph{Computational Science --- ICCS 2002}},
  \bibfield{editor}{\bibinfo{person}{Peter M.~A. Sloot},
  \bibinfo{person}{Alfons~G. Hoekstra}, \bibinfo{person}{C.~J.~Kenneth Tan},
  {and} \bibinfo{person}{Jack~J. Dongarra}} (Eds.).
  \bibinfo{publisher}{Springer}, \bibinfo{address}{Berlin, Heidelberg,
  Germany}, \bibinfo{pages}{1019--1028}.
\newblock
\urldef\tempurl%
\url{https://doi.org/10.1007/3-540-46080-2_107}
\showDOI{\tempurl}


\bibitem[Heimbach et~al\mbox{.}(2005)]%
        {HeimbachHG2005}
\bibfield{author}{\bibinfo{person}{Patrick Heimbach}, \bibinfo{person}{Chris
  Hill}, {and} \bibinfo{person}{Ralf Giering}.}
  \bibinfo{year}{2005}\natexlab{}.
\newblock \showarticletitle{An efficient exact adjoint of the parallel {MIT
  General Circulation Model}, generated via automatic differentiation}.
\newblock \bibinfo{journal}{\emph{Future Gener. Comput. Syst.}}
  \bibinfo{volume}{21}, \bibinfo{number}{8} (\bibinfo{date}{Oct.}
  \bibinfo{year}{2005}), \bibinfo{pages}{1356--1371}.
\newblock
\urldef\tempurl%
\url{https://doi.org/10.1016/j.future.2004.11.010}
\showDOI{\tempurl}


\bibitem[Hogan(2014)]%
        {Hogan2014}
\bibfield{author}{\bibinfo{person}{Robin~J. Hogan}.}
  \bibinfo{year}{2014}\natexlab{}.
\newblock \showarticletitle{Fast reverse-mode automatic differentiation using
  expression templates in {C++}}.
\newblock \bibinfo{journal}{\emph{ACM Trans. Math. Softw.}}
  \bibinfo{volume}{40}, \bibinfo{number}{4}, Article \bibinfo{articleno}{26}
  (\bibinfo{date}{June} \bibinfo{year}{2014}), \bibinfo{numpages}{16}~pages.
\newblock
\urldef\tempurl%
\url{https://doi.org/10.1145/2560359}
\showDOI{\tempurl}


\bibitem[H{\"u}ckelheim and Hasco{\"e}t(2021)]%
        {HueckelheimH2021}
\bibfield{author}{\bibinfo{person}{Jan H{\"u}ckelheim} {and}
  \bibinfo{person}{Laurent Hasco{\"e}t}.} \bibinfo{year}{2021}\natexlab{}.
\newblock \bibinfo{title}{Source-to-source automatic differentiation of
  {OpenMP} parallel loops}.
\newblock
\newblock
\urldef\tempurl%
\url{https://arxiv.org/abs/2111.01861}
\showURL{%
\tempurl}
\newblock
\shownote{Preprint arXiv:2111.01861}.


\bibitem[H{\"u}ckelheim et~al\mbox{.}(2019)]%
        {HueckelheimHSM2019}
\bibfield{author}{\bibinfo{person}{Jan H{\"u}ckelheim}, \bibinfo{person}{Paul
  Hovland}, \bibinfo{person}{Michelle~Mills Strout}, {and}
  \bibinfo{person}{Jens-Dominik M{\"u}ller}.} \bibinfo{year}{2019}\natexlab{}.
\newblock \showarticletitle{Reverse-mode algorithmic differentiation of an
  {OpenMP}-parallel compressible flow solver}.
\newblock \bibinfo{journal}{\emph{Int. J. High Perform. Comput. Appl.}}
  \bibinfo{volume}{33}, \bibinfo{number}{1} (\bibinfo{date}{June}
  \bibinfo{year}{2019}), \bibinfo{pages}{140--154}.
\newblock
\urldef\tempurl%
\url{https://doi.org/10.1177/1094342017712060}
\showDOI{\tempurl}


\bibitem[Kaler et~al\mbox{.}(2021)]%
        {KalerSXLCPK2021}
\bibfield{author}{\bibinfo{person}{Tim Kaler}, \bibinfo{person}{Tao~B.
  Schardl}, \bibinfo{person}{Brian Xie}, \bibinfo{person}{Charles~E.
  Leiserson}, \bibinfo{person}{Jie Chen}, \bibinfo{person}{Aldo Pareja}, {and}
  \bibinfo{person}{Georgios Kollias}.} \bibinfo{year}{2021}\natexlab{}.
\newblock \showarticletitle{{PARAD}: a work-efficient parallel algorithm for
  reverse-mode automatic differentiation}. In
  \bibinfo{booktitle}{\emph{Symposium on Algorithmic Principles of Computer
  Systems (APOCS)}}. \bibinfo{publisher}{SIAM}, \bibinfo{address}{Philadelphia,
  PA, USA}, \bibinfo{pages}{144--158}.
\newblock
\urldef\tempurl%
\url{https://doi.org/10.1137/1.9781611976489.11}
\showDOI{\tempurl}


\bibitem[Kowarz(2008)]%
        {Kowarz2008}
\bibfield{author}{\bibinfo{person}{Andreas Kowarz}.}
  \bibinfo{year}{2008}\natexlab{}.
\newblock \emph{\bibinfo{title}{Advanced concepts for automatic differentiation
  based on operator overloading}}.
\newblock \bibinfo{thesistype}{Ph.\,D. Dissertation}. \bibinfo{school}{Techn.
  Univ. Dresden}, \bibinfo{address}{Dresden, Germany}.
\newblock
\urldef\tempurl%
\url{https://nbn-resolving.org/urn:nbn:de:bsz:14-ds-1206719130404-22306}
\showURL{%
\tempurl}


\bibitem[Nemili et~al\mbox{.}(2016)]%
        {NemiliOGKT2016}
\bibfield{author}{\bibinfo{person}{Anil Nemili}, \bibinfo{person}{Emre
  {\"O}zkaya}, \bibinfo{person}{Nicolas~R. Gauger}, \bibinfo{person}{Felix
  Kramer}, {and} \bibinfo{person}{Frank Thiele}.}
  \bibinfo{year}{2016}\natexlab{}.
\newblock \showarticletitle{Discrete adjoint based optimal active control of
  separation on a realistic high-lift configuration}. In
  \bibinfo{booktitle}{\emph{New Results in Numerical and Experimental Fluid
  Mechanics X}}, \bibfield{editor}{\bibinfo{person}{Andreas Dillmann},
  \bibinfo{person}{Gerd Heller}, \bibinfo{person}{Ewald Kr{\"a}mer},
  \bibinfo{person}{Claus Wagner}, {and} \bibinfo{person}{Christian
  Breitsamter}} (Eds.). \bibinfo{publisher}{Springer}, \bibinfo{address}{Cham,
  Switzerland}, \bibinfo{pages}{237--246}.
\newblock
\urldef\tempurl%
\url{https://doi.org/10.1007/978-3-319-27279-5_21}
\showDOI{\tempurl}


\bibitem[{OpenMP Architecture Review Board}(2005)]%
        {OpenMP25}
\bibfield{author}{\bibinfo{person}{{OpenMP Architecture Review Board}}.}
  \bibinfo{year}{2005}\natexlab{}.
\newblock \bibinfo{title}{{OpenMP Application Program Interface Version 2.5}}.
\newblock
\newblock
\urldef\tempurl%
\url{https://www.openmp.org/wp-content/uploads/spec25.pdf}
\showURL{%
\tempurl}


\bibitem[{OpenMP Architecture Review Board}(2013)]%
        {OpenMP40}
\bibfield{author}{\bibinfo{person}{{OpenMP Architecture Review Board}}.}
  \bibinfo{year}{2013}\natexlab{}.
\newblock \bibinfo{title}{{OpenMP Application Program Interface Version 4.0}}.
\newblock
\newblock
\urldef\tempurl%
\url{https://www.openmp.org/wp-content/uploads/OpenMP4.0.0.pdf}
\showURL{%
\tempurl}


\bibitem[{OpenMP Architecture Review Board}(2018)]%
        {OpenMP50}
\bibfield{author}{\bibinfo{person}{{OpenMP Architecture Review Board}}.}
  \bibinfo{year}{2018}\natexlab{}.
\newblock \bibinfo{title}{{OpenMP Application Programming Interface Version
  5.0}}.
\newblock
\newblock
\urldef\tempurl%
\url{https://www.openmp.org/wp-content/uploads/OpenMP-API-Specification-5.0.pdf}
\showURL{%
\tempurl}


\bibitem[{OpenMP Architecture Review Board}(2021)]%
        {OpenMP52}
\bibfield{author}{\bibinfo{person}{{OpenMP Architecture Review Board}}.}
  \bibinfo{year}{2021}\natexlab{}.
\newblock \bibinfo{title}{{OpenMP Application Programming Interface Version
  5.2}}.
\newblock
\newblock
\urldef\tempurl%
\url{https://www.openmp.org/wp-content/uploads/OpenMP-API-Specification-5-2.pdf}
\showURL{%
\tempurl}


\bibitem[{\"O}zkaya and Gauger(2010)]%
        {OezkayaG2010}
\bibfield{author}{\bibinfo{person}{Emre {\"O}zkaya} {and}
  \bibinfo{person}{Nicolas~R. Gauger}.} \bibinfo{year}{2010}\natexlab{}.
\newblock \showarticletitle{Automatic transition from simulation to one-shot
  shape optimization with {N}avier-{S}tokes equations}.
\newblock \bibinfo{journal}{\emph{GAMM-Mitteilungen}} \bibinfo{volume}{33},
  \bibinfo{number}{2} (\bibinfo{date}{Oct.} \bibinfo{year}{2010}),
  \bibinfo{pages}{133--147}.
\newblock
\urldef\tempurl%
\url{https://doi.org/10.1002/gamm.201010011}
\showDOI{\tempurl}


\bibitem[Protze et~al\mbox{.}(2019)]%
        {ProtzeCCM2019}
\bibfield{author}{\bibinfo{person}{Joachim Protze}, \bibinfo{person}{Tim
  Cramer}, \bibinfo{person}{Simon Convent}, {and} \bibinfo{person}{Matthias~S.
  M{\"u}ller}.} \bibinfo{year}{2019}\natexlab{}.
\newblock \showarticletitle{{OMPT-Multiplex}: nesting of {OMPT} tools}. In
  \bibinfo{booktitle}{\emph{Tools for High Performance Computing 2017}},
  \bibfield{editor}{\bibinfo{person}{Christoph Niethammer},
  \bibinfo{person}{Michael~M. Resch}, \bibinfo{person}{Wolfgang~E. Nagel},
  \bibinfo{person}{Holger Brunst}, {and} \bibinfo{person}{Hartmut Mix}} (Eds.).
  \bibinfo{publisher}{Springer}, \bibinfo{address}{Cham, Switzerland},
  \bibinfo{pages}{73--83}.
\newblock
\urldef\tempurl%
\url{https://doi.org/10.1007/978-3-030-11987-4_5}
\showDOI{\tempurl}


\bibitem[Sagebaum et~al\mbox{.}(2018)]%
        {SagebaumAG2018}
\bibfield{author}{\bibinfo{person}{M. Sagebaum}, \bibinfo{person}{T. Albring},
  {and} \bibinfo{person}{N.~R. Gauger}.} \bibinfo{year}{2018}\natexlab{}.
\newblock \showarticletitle{Expression templates for primal value taping in the
  reverse mode of algorithmic differentiation}.
\newblock \bibinfo{journal}{\emph{Optim. Methods Software}}
  \bibinfo{volume}{33}, \bibinfo{number}{4-6} (\bibinfo{date}{May}
  \bibinfo{year}{2018}), \bibinfo{pages}{1207--1231}.
\newblock
\urldef\tempurl%
\url{https://doi.org/10.1080/10556788.2018.1471140}
\showDOI{\tempurl}


\bibitem[Sagebaum et~al\mbox{.}(2019)]%
        {SagebaumAG2019}
\bibfield{author}{\bibinfo{person}{Max Sagebaum}, \bibinfo{person}{Tim
  Albring}, {and} \bibinfo{person}{Nicolas~R. Gauger}.}
  \bibinfo{year}{2019}\natexlab{}.
\newblock \showarticletitle{High-performance derivative computations using
  {CoDiPack}}.
\newblock \bibinfo{journal}{\emph{ACM Trans. Math. Softw.}}
  \bibinfo{volume}{45}, \bibinfo{number}{4}, Article \bibinfo{articleno}{34}
  (\bibinfo{date}{Dec.} \bibinfo{year}{2019}), \bibinfo{numpages}{26}~pages.
\newblock
\urldef\tempurl%
\url{https://doi.org/10.1145/3356900}
\showDOI{\tempurl}


\bibitem[Sagebaum et~al\mbox{.}(2020)]%
        {SagebaumBG2020}
\bibfield{author}{\bibinfo{person}{Max Sagebaum}, \bibinfo{person}{Johannes
  Bl{\"u}hdorn}, {and} \bibinfo{person}{Nicolas~R. Gauger}.}
  \bibinfo{year}{2020}\natexlab{}.
\newblock \bibinfo{title}{Index handling and assign optimization for
  algorithmic differentiation reuse index managers}.
\newblock
\newblock
\urldef\tempurl%
\url{https://arxiv.org/abs/2006.12992}
\showURL{%
\tempurl}
\newblock
\shownote{Preprint arXiv:2006.12992}.


\bibitem[Sagebaum et~al\mbox{.}(2013)]%
        {SagebaumGNLL2013}
\bibfield{author}{\bibinfo{person}{Max Sagebaum}, \bibinfo{person}{Nicolas~R.
  Gauger}, \bibinfo{person}{Uwe Naumann}, \bibinfo{person}{Johannes Lotz},
  {and} \bibinfo{person}{Klaus Leppkes}.} \bibinfo{year}{2013}\natexlab{}.
\newblock \showarticletitle{Algorithmic differentiation of a complex {C++} code
  with underlying libraries}, In \bibinfo{booktitle}{2013 International
  Conference on Computational Science},
  \bibfield{editor}{\bibinfo{person}{Vassil Alexandrov},
  \bibinfo{person}{Michael Lees}, \bibinfo{person}{Valeria Krzhizhanovskaya},
  \bibinfo{person}{Jack Dongarra}, {and} \bibinfo{person}{Peter M.~A. Sloot}}
  (Eds.).
\newblock \bibinfo{journal}{\emph{Procedia Comput. Sci.}}
  \bibinfo{volume}{18}, \bibinfo{pages}{208--217}.
\newblock
\showISSN{1877-0509}
\urldef\tempurl%
\url{https://doi.org/10.1016/j.procs.2013.05.184}
\showDOI{\tempurl}


\bibitem[Schanen et~al\mbox{.}(2012)]%
        {SchanenFLLN2012}
\bibfield{author}{\bibinfo{person}{Michel Schanen}, \bibinfo{person}{Michael
  F{\"o}rster}, \bibinfo{person}{Johannes Lotz}, \bibinfo{person}{Klaus
  Leppkes}, {and} \bibinfo{person}{Uwe Naumann}.}
  \bibinfo{year}{2012}\natexlab{}.
\newblock \showarticletitle{Adjoining hybrid parallel code}. In
  \bibinfo{booktitle}{\emph{Proceedings of the Eighth International Conference
  on Engineering Computational Technology}} \emph{(\bibinfo{series}{Civil-Comp
  Proceedings}, Vol.~\bibinfo{volume}{100})},
  \bibfield{editor}{\bibinfo{person}{B.~H.~V. Topping}} (Ed.).
  \bibinfo{publisher}{Civil-Comp Press}, \bibinfo{address}{Kippen,
  Stirlingshire, UK}, \bibinfo{numpages}{18}~pages.
\newblock
\urldef\tempurl%
\url{https://doi.org/10.4203/ccp.100.7}
\showDOI{\tempurl}


\bibitem[Schanen et~al\mbox{.}(2010)]%
        {SchanenNHU2010}
\bibfield{author}{\bibinfo{person}{Michel Schanen}, \bibinfo{person}{Uwe
  Naumann}, \bibinfo{person}{Laurent Hasco{\"e}t}, {and} \bibinfo{person}{Jean
  Utke}.} \bibinfo{year}{2010}\natexlab{}.
\newblock \showarticletitle{Interpretative adjoints for numerical simulation
  codes using {MPI}}, In \bibinfo{booktitle}{2010 International Conference on
  Computational Science}.
\newblock \bibinfo{journal}{\emph{Procedia Comput. Sci.}} \bibinfo{volume}{1},
  \bibinfo{number}{1}, \bibinfo{pages}{1825--1833}.
\newblock
\urldef\tempurl%
\url{https://doi.org/10.1016/j.procs.2010.04.204}
\showDOI{\tempurl}


\bibitem[Schlenkrich et~al\mbox{.}(2008)]%
        {SchlenkrichWGH2008}
\bibfield{author}{\bibinfo{person}{Sebastian Schlenkrich},
  \bibinfo{person}{Andrea Walther}, \bibinfo{person}{Nicolas~R. Gauger}, {and}
  \bibinfo{person}{Ralf Heinrich}.} \bibinfo{year}{2008}\natexlab{}.
\newblock \showarticletitle{Differentiating fixed point iterations with
  {ADOL-C}: gradient calculation for fluid dynamics}. In
  \bibinfo{booktitle}{\emph{Modeling, Simulation and Optimization of Complex
  Processes}}, \bibfield{editor}{\bibinfo{person}{Hans~Georg Bock},
  \bibinfo{person}{Ekaterina Kostina}, \bibinfo{person}{Hoang~Xuan Phu}, {and}
  \bibinfo{person}{Rolf Rannacher}} (Eds.). \bibinfo{publisher}{Springer},
  \bibinfo{address}{Berlin, Heidelberg, Germany}, \bibinfo{pages}{499--508}.
\newblock
\urldef\tempurl%
\url{https://doi.org/10.1007/978-3-540-79409-7_36}
\showDOI{\tempurl}


\bibitem[Towara and Naumann(2013)]%
        {TowaraN2013}
\bibfield{author}{\bibinfo{person}{Markus Towara} {and} \bibinfo{person}{Uwe
  Naumann}.} \bibinfo{year}{2013}\natexlab{}.
\newblock \showarticletitle{A discrete adjoint model for {OpenFOAM}}, In
  \bibinfo{booktitle}{2013 International Conference on Computational Science},
  \bibfield{editor}{\bibinfo{person}{Vassil Alexandrov},
  \bibinfo{person}{Michael Lees}, \bibinfo{person}{Valeria Krzhizhanovskaya},
  \bibinfo{person}{Jack Dongarra}, {and} \bibinfo{person}{Peter M.~A. Sloot}}
  (Eds.).
\newblock \bibinfo{journal}{\emph{Procedia Comput. Sci.}}
  \bibinfo{volume}{18}, \bibinfo{pages}{429--438}.
\newblock
\urldef\tempurl%
\url{https://doi.org/10.1016/j.procs.2013.05.206}
\showDOI{\tempurl}


\bibitem[Verstraete et~al\mbox{.}(2017)]%
        {VerstraeteMM2017}
\bibfield{author}{\bibinfo{person}{Tom Verstraete}, \bibinfo{person}{Lasse
  Müller}, {and} \bibinfo{person}{Jens-Dominik Müller}.}
  \bibinfo{year}{2017}\natexlab{}.
\newblock \showarticletitle{Adjoint-based design optimisation of an internal
  cooling channel u-bend for minimised pressure losses}.
\newblock \bibinfo{journal}{\emph{Int. J. Turbomach. Propuls. Power}}
  \bibinfo{volume}{2}, \bibinfo{number}{2} (\bibinfo{date}{May}
  \bibinfo{year}{2017}), \bibinfo{numpages}{13}~pages.
\newblock
\urldef\tempurl%
\url{https://doi.org/10.3390/ijtpp2020010}
\showDOI{\tempurl}


\bibitem[Vitale et~al\mbox{.}(2020)]%
        {VitalePC2020}
\bibfield{author}{\bibinfo{person}{S. Vitale}, \bibinfo{person}{M. Pini}, {and}
  \bibinfo{person}{P. Colonna}.} \bibinfo{year}{2020}\natexlab{}.
\newblock \showarticletitle{Multistage turbomachinery design using the discrete
  adjoint method within the open-source software {SU2}}.
\newblock \bibinfo{journal}{\emph{Journal Propul. Power}} \bibinfo{volume}{36},
  \bibinfo{number}{3} (\bibinfo{date}{Mar.} \bibinfo{year}{2020}),
  \bibinfo{pages}{465--478}.
\newblock
\urldef\tempurl%
\url{https://doi.org/10.2514/1.B37685}
\showDOI{\tempurl}


\bibitem[Walther(2009)]%
        {Walther2009}
\bibfield{author}{\bibinfo{person}{Andrea Walther}.}
  \bibinfo{year}{2009}\natexlab{}.
\newblock \showarticletitle{Getting started with {ADOL-C}}. In
  \bibinfo{booktitle}{\emph{Combinatorial Scientific Computing}}
  \emph{(\bibinfo{series}{Dagstuhl Seminar Proceedings},
  Vol.~\bibinfo{volume}{9061})}, \bibfield{editor}{\bibinfo{person}{Uwe
  Naumann}, \bibinfo{person}{Olaf Schenk}, \bibinfo{person}{Horst~D. Simon},
  {and} \bibinfo{person}{Sivan Toledo}} (Eds.). \bibinfo{publisher}{Schloss
  Dagstuhl -- Leibniz-Zentrum f{\"u}r Informatik}, \bibinfo{address}{Dagstuhl,
  Germany}, \bibinfo{numpages}{10}~pages.
\newblock
\urldef\tempurl%
\url{https://drops.dagstuhl.de/opus/volltexte/2009/2084}
\showURL{%
\tempurl}


\bibitem[Zhu et~al\mbox{.}(2010)]%
        {Zhu2010}
\bibfield{author}{\bibinfo{person}{Hongqing Zhu}, \bibinfo{person}{Huazhong
  Shu}, {and} \bibinfo{person}{Meiyu Ding}.} \bibinfo{year}{2010}\natexlab{}.
\newblock \showarticletitle{Numerical solutions of two-dimensional {B}urgers'
  equations by discrete {A}domian decomposition method}.
\newblock \bibinfo{journal}{\emph{Comput. Math. Appl.}} \bibinfo{volume}{60},
  \bibinfo{number}{3} (\bibinfo{date}{Aug.} \bibinfo{year}{2010}),
  \bibinfo{pages}{840--848}.
\newblock
\urldef\tempurl%
\url{https://doi.org/10.1016/j.camwa.2010.05.031}
\showDOI{\tempurl}


\end{thebibliography}

\end{document}